\documentclass{natureprintstylev4}
\usepackage{astjnlabbrev-nature}
\usepackage{hyperref}
\usepackage[switch]{lineno}

\usepackage{ulem}
\usepackage{array}
\usepackage{amssymb}
\usepackage{mathptmx}
\usepackage{amsmath}	
\usepackage{gensymb}
\usepackage{color}
\usepackage{enumitem}
\usepackage{textcomp}
\usepackage{marvosym}

\usepackage{epsfig}
\usepackage{color}
\usepackage{graphicx}
\usepackage{longtable}
\usepackage{hyperref}
\usepackage{float}

\usepackage[switch]{lineno}

\usepackage[labelsep=endash]{caption}

\newcommand{\citep}{\cite}
\newcommand{\citet}{\cite}
\usepackage[utf8]{inputenc}
\title{A slightly oblate dark matter halo revealed by a retrograde precessing Galactic disk warp}
\author{Yang Huang$^{1,2, \dagger, \textsuperscript{\Letter}}$, Qikang Feng$^{3,4, \dagger}$, Tigran Khachaturyants$^{5,6}$, Huawei Zhang$^{3,4, \textsuperscript{\Letter}}$, Jifeng Liu$^{2,1,7,8,\textsuperscript{\Letter}}$, Juntai Shen$^{5,6,\textsuperscript{\Letter}}$, Timothy C. Beers$^9$, Youjun Lu$^{7,1}$, Song Wang$^{7,2,8}$, Haibo Yuan$^{8,10}$}

\begin{document}
\maketitle
\begin{affiliations}
\item School of Astronomy and Space Science, University of Chinese Academy of Sciences, Beijing 100049, People's Republic of China;\\\
\item  New Cornerstone Science Laboratory, National Astronomical Observatories, Chinese Academy of Sciences, Beijing 100012, People’s Republic of China;\\
\item Department of Astronomy, School of Physics, Peking University, Beijing 100871, People's Republic of China;\\
\item Kavli Institute for Astronomy and Astrophysics, Peking University, Beijing 100871, People's Republic of China;\\
\item Department of Astronomy, School of Physics and Astronomy, Shanghai Jiao Tong University, Shanghai 200240, People’s Republic of China;\\
\item Key Laboratory for Particle Astrophysics and Cosmology (MOE)/Shanghai Key Laboratory for Particle Physics and Cosmology, Shanghai 200240, People’s Republic of China\\
\item National Astronomical Observatories, Chinese Academy of Sciences, Beijing 100012, People’s Republic of China;\\
\item Institute for Frontiers in Astronomy and Astrophysics, Beijing Normal University, Beijing, 102206, People’s Republic of China;\\
\item Department of Physics and Astronomy and JINA Center for the Evolution of the Elements (JINA-CEE), University of Notre Dame, Notre Dame, IN 46556, USA;\\
\item Department of Astronomy, Beijing Normal University, Beijing, 100875, People's Republic of China;\\
$\dagger$ {These authors contributed equally};\\
$\textsuperscript{\Letter}$ Corresponding authors:  jfliu@nao.cas.cn; huangyang@ucas.ac.cn;\
zhanghw@pku.edu.cn; jtshen@sjtu.edu.cn.\\
\end{affiliations}
\\

\section*{Abstract}
\textbf{The shape of the dark matter (DM) halo is key to understanding the hierarchical formation of the Galaxy. 
Despite extensive efforts in the recent decades, 
however, its shape remains a matter of debate, with suggestions ranging from strongly oblate to prolate.
Here, we present a new constraint on its present shape by directly measuring the evolution of the Galactic disk warp with time, as traced by accurate distance estimates and precise age determinations for about 2,600 classical Cepheids.
We show that the Galactic warp is mildly precessing in a retrograde direction at a rate of $\omega = -2.1 \pm 0.5\,({\rm statistical}) \pm 0.6\,({\rm systematic})$\,km\,s$^{-1}$\,kpc$^{-1}$ for the outer disk over the Galactocentric radius  [$7.5, 25$]\,kpc, decreasing with radius.
This constrains the shape of the DM halo 
to be slightly oblate with a flattening (minor-to-major axis ratio) in the range of $0.84 \le q_{\Phi} \le 0.96$.
Given the young nature of the disk warp traced by Cepheids (less than 200 Myr), our approach directly measures the shape of the present-day DM halo. 
This measurement, combined with other measurements from older tracers, could provide vital constraints on the evolution of the DM halo and the assembly history of the Galaxy.}

A disk warp is a ubiquitous large-scale feature of disk galaxies, including our own\citep{Kerr57}$^{\text{-}}$\citep{Binney92}. 
Theoretically, these warps are the response of the disk to external torques from a variety of sources, including cosmic infall\citep{SS06}$^{-}$\citep{Jeon09}, misalignment between a non-spherical DM halo and the disk\citep{SC88}${^{,}}$\citep{DC09}, interactions with satellite galaxies\citep{WB06}, and the inter-galactic magnetic field \citep{BF90}. Amongst them, the torque exerted by the DM halo plays a major role\citep{Binney92}. This latter torque can be probed by the precession of the warp.
The Galactic disk warp has long been expected to precess in the retrograde direction\citep{SS06}$^{,}$\citep{SC88}$^{,}$\citep{JB99}, i.e., opposite to the Solar rotational motion. Using an indirect kinematic approach, recent efforts have unexpectedly found a precession in the prograde direction\citep{Poggio20}${^,}$\citep{Cheng20}. 
However, these measurements used old giant stars as stellar tracers. Such stars may suffer from complex heating and perturbation histories that could invalidate the results. Furthermore, that approach, which depends on mapping the vertical motion patterns of stellar tracers at different disk locations, usually has rather large uncertainties\citep{Chrob21}, even with young tracers such as Cepheids\citep{Dehnen23} (Methods).

In this study, we develop a ``motion-picture" technique to trace the changing orientation of the disk warp using stellar tracers of different ages, and we compute its precession rate directly by examining its line of nodes (LON) at different times. For stellar tracers, we analyzed classical Cepheids, which are relatively recently born compared to giant stars, as their distances and ages can be measured well by calibrated period--luminosity and period--age--metallicity (PAZ) relations. The data for classical Cepheids used here are taken from the newly released Gaia Data Release\,3 (hereafter Gaia DR3)\citep{Gaia22}$^{,}$\citep{Ripepi22}. 
The distances to these classical Cepheids were precisely determined using the period--Wesenheit (PW) relations\citep{Ripepi22} (Methods).
A comparison with the Cepheids in open clusters (OCs) showed that the PW distances are in excellent agreement with those determined by parallax measurements or isochrone fitting\citep{Hao22}, with a negligible offset of about 1.4\% and a small scatter of 6.8\%.
Further, we removed Cepheids in high-extinction regions, those with distance errors larger than 6\%, and other significant outliers,
leading to a final sample of 2,613 classical Cepheids.
As shown in Figure~1 (a-c), the Galatic disk warp is clearly present in the spatial distributions of the full sample and subsamples with different ages.
\par
Age is the key to measuring the precession rate of the disk warp. The age of Cepheids can be precisely determined by the PAZ relation\citep{DeSomma21}. 
However, only about one-third of our sample of stars have metallicity estimates ([Fe/H]) in the Gaia Cepheid catalog.
For the remaining two-thirds, their metallicities were estimated based on their Galactocentric radius $R$ by adopting a Galactic radial metallicity distribution (Methods). 
In this way, age estimates are derived for all sample stars with a precision better than 20\%.
The derived ages for Cepheids in OCs agree very well with ages determined by isochrone fitting\citep{Zhou21}, with a scatter of about 18\%, consistent with the expected uncertainty of this relation (Methods). 
The median age of the full sample is about 100\,Myr; 85\% of them are younger than 200\,Myr (Extended Data Figure\,1).
These Cepheid stars are sufficiently young that they do not experience complex heating or have perturbation histories, in contrast to the much older giant stars. Thus, they retain information about the shape of the warp at the time of their birth.

\par
We obtained a motion picture of the disk warp by mapping the three-dimensional distributions for Cepheid samples of different ages.
We adopted a canonical model to describe the shape of the warp\citep{Poggio17}:
 \begin{equation}
     Z=
     \begin{cases}
     c(R-R_{\rm s})^\alpha sin(\phi-\phi_w), & R > R_{\rm s},\\
     0, & R \le R_{\rm s}.
     \end{cases}
 \end{equation}
In this model, the vertical displacement $Z$ from the disk plane increases as a power law with an index of $\alpha$, varies in a sinusoidal fashion with respect to the Galactic azimuth $\phi$ and begins to warp at $R_{\rm s}$.
The $\phi_w$ parameter is the phase angle of the LON along which the vertical displacement is zero.
As a first step, this model was fitted to the full Cepheid sample, which includes stars of all ages, using a least-squares algorithm.
We verified that the fitting is insensitive to the warp-starting radius $R_{\rm s}$, so it was fixed to $7.5$\,kpc after careful checks (Methods).
The best-fitting model yields a power-law index of $1.40 \pm 0.05$ and a LON with $\phi_w=10.06\pm 0.93^{\circ}$. These geometric parameters are consistent with previous measurements from various tracers\citep{Chen19}$^{,}$\citep{Burton88}$^{,}$\citep{Li20}.
As shown in Figure\,1(a) and Extended Data Figure\,2, such a warp model describes the spatial distribution of Cepheids quite well.
\par
For an evolving disk warp, $\phi_w$ is a linear function of cosmic time $t$: $\phi_w (t) = \phi_{0, w} + \omega (t - t_0)$.
Here, $\phi_{0, w}$ represents the current LON of the Galactic warp, $\omega$ is the precession rate, $t_0$ is the cosmic age of the universe and $t - t_0 = -\langle \tau \rangle$, where $\langle \tau \rangle$ is the median age of the stellar sample.
To derive $\omega$, we measured $\phi_w$ for Cepheid subsamples of different ages.
The sample stars were divided into eight age bins of width 100 Myr (Table\,1).
The bins overlap each other with a running step of 20\,Myr to ensure there are sufficient numbers of Cepheids in each bin.
The median ages of these bins range from 80\,Myr to 160\,Myr, with an entire age span of about 80\,Myr.
As in the analysis for the full sample, the warp model was then fitted to Cepheid 
subsamples of different ages.
The fitting results are presented in Table\,1 and shown in Figure\,1(d).
As the Cepheid subsamples become younger, the LON $\phi_w$ tends to become smaller, which means that the disk warp is precessing in a retrograde fashion,  as long expected\citep{SS06}$^{,}$\citep{SC88}$^{,}$\citep{JB99}.

To quantitatively measure the precession rate $\omega$, a linear fit was applied to the measured $\phi_w (t)$ as shown in 
Figure\,1(d). Given the wide range of the age bins, Deming regression was employed to account for uncertainties in both age and LON during the linear fitting. This analysis yielded a mildly retrograde precession rate of $\omega = -2.1 \pm 0.5$\,km\,s\,$^{-1}$\,kpc$^{-1}$ (equal to $0.12 \pm 0.03^{\circ}$\,Myr$^{-1}$). 
The systematic error is smaller than 
0.6\,km\,s\,$^{-1}$\,kpc$^{-1}$, which was estimated by considering uncertainties from the choices of different values of $R_{\rm s}$, determinations of Cepheid distances and ages, and potential selection effects in the tracer sample (Methods). 
In contrast to this result, a large prograde precession rate of $\omega = 10.86 \pm 0.03 \ ({\rm statistical}) \pm 3.20\ ({\rm systematic})$\,km\,s\,$^{-1}$\,kpc$^{-1}$ was found from an analysis of old giant stars based on a kinematic approach\citep{Poggio20}.
To verify our results, we remeasured the precession rate of the disk warp based on the kinematic approach, but using young stellar tracers, namely around 1,200 Cepheids with high-quality radial velocity measurements from Gaia DR3 (Methods).
The resulting $\omega = -1.1 \pm 1.9$\,km\,s\,$^{-1}$\,kpc$^{-1}$ is consistent with our own measurement but with uncertainties much larger than those obtained through our motion-picture approach.
\par
Our more accurate measurement for the warp's precession rate offers a unique opportunity to constrain the shape of the DM halo. We adopted a simple model to calculate the precession rate at different radii analytically, with major contributions from the Galactic disk and the DM halo (Methods).
The former can be directly derived, as the structural parameters and total mass of the disk are relatively well measured, whereas the latter is highly dependent on the shape of the DM halo, which is usually characterized by the flattening (minor-to-major axis ratio) $q_{\Phi}$. Note that the shape of the DM halo may be non-axisymmetric, and this asymmetry could possibly induce the disk warp\citep{HAN23}.
To constrain $q_{\Phi}$, we further divide the Cepheid sample into three radial bins: $11.8 \le R \le 18.8$\,kpc, $14 \le R \le 21$\,kpc, and $R \ge 15.5$\,kpc. The choice of three bins was a trade-off between the number of stars and having sufficient range to detect a clear signal of the warp and its precession.
The procedure to measure the warp-precession rate was applied to the three bins. The results show a decreasing trend with $R$ (Figure\,2 (a)).
By subtracting the contributions from disks, the residual precession rates are clearly retrograde, with values ranging from $-1.5$ to $-1.0$\ \,km\,s\,$^{-1}$\,kpc$^{-1}$ (Extended Data Figure 3), suggesting that the DM halo is oblate  rather than spherical or prolate.
By comparing the measured precession rates with our toy model, the flattening of the DM halo was found to be in the range of $0.84 \le q_{\Phi} \le 0.96$ (Methods).

Our measurement of $q_{\Phi}$ from the disk-warp precession revealed that the DM halo is slightly oblate.  
This result is largely consistent with measurements from stellar-stream analysis within errors\citep{SB13}$^{,}$\citep{Bovy16} $^{,}$\citep{Bowden15}$^{,}$\citep{K15}, but is inconsistent with measurements based on halo stars\citep{Loebman14}$^{,}$\citep{Wegg19} or globular clusters\citep{PH19}. 
Such inconsistencies may be caused by unaccounted-for systematic errors in the different methods or may reflect the intrinsic evolution of DM halo shape itself\citep{Cataldi23}, since different measurements are sensitive to torques at different cosmic times or intervals of times.
Our measurement probes the present-day shape (in the past 200 Myrs), and provides an anchoring point across cosmic history; if other measurements can be accurately 
time-tagged in future studies, the evolution of the DM halo shape can be fully revealed, which may shed light on the assembly history of the Galaxy.
 
\section*{Methods}
\subsection*{Coordinate systems.}
In this study, two sets of coordinate systems are adopted: 
(1) a right-handed Galactocentric Cartesian coordinate system ($X$, $Y$, $Z$), with positive $X$ direction pointing towards the Galactic centre from the Sun, $Y$ towards the direction of Galactic rotation of the Sun, and $Z$ in the direction of the north Galactic pole;
(2) a Galactocentric cylindrical system ($R$, $\phi$, $Z$), with $R$ increasing radially outwards, $\phi$ the azimuthal angle pointing in the direction of Galactic rotation, and $Z$ the same as that in the Cartesian system.
The Sun is fixed at ($-8.178$, 0, 0.025)\,kpc in Cartesian coordinates\citep{GRAVITY19}$^{,}$\citep{Juri08}.
The Galactocentric velocity of the Sun is fixed to $V_{R, \odot}=11.1\ {\rm km}\ {\rm s}^{-1}$ (ref.\citep{Sch10}), $V_{\phi, \odot}=245.6\ {\rm km}\ {\rm s}^{-1}$ and $V_{Z, \odot}=7.8\ {\rm km}\ {\rm s}^{-1}$(ref.\citep{Reid04}).

\subsection*{Cepheid sample, distance estimates, and validations.}
\par
Our Cepheid sample was from the Gaia DR3 Cepheid catalog\citep{Ripepi22}, which was downloaded via the Gaia Archive (\url{https://gea.esac.esa.int/archive/}).
Based on multi-band time-series photometry, Gaia DR3 contains a sample of 15,006 Cepheids of all types, as yielded by the SOS Cep\&RRL pipeline\citep{Clementini2022}.
We selected only the Milky Way classical Cepheids (labelled as `DCEP' in the catalog) with a Gaia renormalized unit weight error $<1.4$, 
which guards against poor astrometry.
The completeness of this sample is greater than 85\%, and the contamination is at the level of only a few per cent.
\par
To reduce the uncertainties of extinction corrections, the PW relations were adopted to derive the distances of the Cepheids, thanks to the accurate multi-band time-series photometry provided by Gaia.
The Wesenheit magnitude is defined to be extinction free and here is expressed as: $w = G - k(G_{\rm BP} - G_{\rm RP})$, where $k = \frac{A_G}{E(G_{\rm BP} - G_{\rm RP})} = 1.90$ (ref.\citep{Ripepi22}), and $G$, $G_{\rm BP}$, and $G_{\rm RP}$ are the intensity-averaged magnitudes from the light curves\citep{Ripepi22}. The absolute Wesenheit magnitudes of Cepheids can be predicted by their well-determined periods based on the  PW relations: $W = \alpha + \beta {\rm log}(P)$, for which the values of $\alpha$ and $\beta$ were properly re-calibrated using the Gaia DR3 data for all-sky Cepheids\citep{Ripepi22}.
Finally, the distances of Cepheids were derived with: $d = 10^{0.2(w-W+5)}$.
The distance distribution of the full Cepheid sample is presented in Supplementary Figure\,1. The most distant stars are as far as 25\,kpc from the Sun.
\par
To examine the robustness of distance estimates from the PW relations, we further checked the distances of Cepheids by cross-matching our sample to the compiled catalog of Cepheid--OC pairs\cite{Hao22}.
Using a matching radius of 10 to 15 arcsec, 21 Cepheid--OC pairs were found.
As shown in Supplementary Figure\,2, the distances of these 21 Cepheids yielded by the PW relations are in excellent agreement with 19 nearby OCs ($\overline{\varpi_{\rm OC}} > 0.15$\,mas), as measured from the mean parallaxes of their members. The distances of two distant OCs ($\overline{\varpi_{\rm OC}} \leq 0.15$\,mas) determined by isochrone fitting\citep{Zhou21}$^{,}$\citep{Negueruela18}.
The overall offset of the relative distance differences was 1.4\%, with a scatter of 6.8\%.
This comparison clearly demonstrates the robustness of the distance estimates from application of the PW relations for the classical Cepheids in this study.

\subsection*{Age estimates and validation.}
\par
The ages of the Cepheids were derived from the period-age-metallicity (PAZ) relation, which was properly calibrated based on pulsation models for classical Cepheids\citep{DeSomma21}:
\begin{equation}
\begin{split}
{\rm log} (\tau) = (8.423 \pm 0.006) - (0.642 \pm 0.004){\rm log} P\\
- (0.067 \pm 0.006) {\rm [Fe/H]}\text{.}
\end{split}
\end{equation}
Note that this relation is only valid for fundamental-mode Cepheids.
To derive ages of first-overtone mode Cepheids, their periods were fundamentalized via the empirical relation\citep{FC97}: $P_{\rm F} = P_{\rm 1O}/(0.716 - 0.027{\rm log} P_{\rm 1O})$, where $P_{\rm F}$ and $P_{\rm 1O}$ are, respectively, the periods of the fundamental and first-overtone modes.

To derive the ages of Cepheids, metallicity information is required in addition to the periods. 
In the Gaia Cepheid sample, the metallicity for one-third of the sample stars were properly estimated from their light curves\citep{Ripepi22}.
For the remaining two-thirds, we assigned metallicities according to their radial positions on the disk plane by comparing them with the
radial distribution of [Fe/H] using the one-third of sample stars with known metallicity. 
The radial metallicity distribution, presented in Supplementary Figure\,3, shows a clear negative radial gradient for $R$ within about 16\,kpc and then tends to flatten outside this radius.
This distribution is like the results found with other disk tracers\citep{Hayden14}$^{,}$\citep{Huang15}. 
To quantitatively describe this distribution, the sample stars were divided into radial bins with a width of 1\,kpc for $6 < R < 20$\,kpc.
Over the ranges of $4 < R < 6$\,kpc and $20 < R < 25$\,kpc the radial bins were assigned larger widths to include a sufficient number of stars.
In each radial bin, the values of the median and scatter of the metallicity distribution were calculated after 2$\sigma$ to 3$\sigma$ clipping.
A piece-wise function was applied to this radial trend of the median metallicity:
\begin{equation}
    \rm{[Fe/H]}=
    \begin{cases}
    kR+z_0, & R\le R_{\rm b},\\
    kR_{\rm b}+z_0, & R>R_{\rm b},
    \end{cases}
\end{equation}
where a linear function was used to describe the radial negative metallicity gradient for $R < R_{\rm b}$, and a flat function was used for $R \ge R_{\rm b}$.
The fits are shown in Supplementary Figure\,3 and yield a radial metallicity gradient $k = -0.037 \pm 0.005$\,dex\,kpc$^{-1}$, a break radius $R_{\rm b} = 16.4 \pm 1.2$\,kpc, and an intercept $z_0 = 0.44 \pm 0.05$.
The metallicities for these stars were then assigned by assuming a Gaussian distribution with a mean value derived from Equation\,3 and a dispersion taken from the scatter at its associated radial bin.
The metallicity distribution for the two-thirds of the Cepheids derived from the radial metallicity distribution is shown in Supplementary Figure\,4, which is very closely resembles that for the one-third of the sample stars with known metallicity from the Gaia catalog.

The metallicity estimated for the two-thirds of the sample stars where it was inferred but not measured was associated with a large uncertainty (typically of order 0.18\,dex). However, this had to negligible effects on the age determinations of Cepheids.
The contribution of the metallicity term in the PAZ relation (Equation\,2) had a minor impact on the Cepheid age, only about one-tenth of the contribution of the period term.
Thus, the uncertainty in the age estimate was no more than 10\%, even if the metallicity of a Cepheid star is incorrectly estimated by as much as 0.5\,dex.
\par
With metallicity determined as above, we derived age estimates for all sample stars using Equation\,2.
The final age distribution is presented in Extended Data Figure\,1.
Most of the sample stars are younger than 200\,Myr, with the oldest one no more than 600\,Myr.
To validate the age estimates, Cepheid--OC pairs were again used. We cross-matched our sample stars to two compiled catalogs\citep{Hao22}$^{,}$\citep{Zhou21}, finding 25 Cepheid--OC pairs with ages properly estimated from isochrone fitting.
As shown in Supplementary Figure\,2,  our age estimates of Cepheids are quite consistent with those of OCs determined by isochrone fitting, with a small scatter of 18.4\% (0.08\,dex in log$\tau$).
The systematic difference was about 40\% (0.15\,dex in log$\tau$), which could due to various reasons.  For example, it could easily have been caused by the different stellar-evolution models adopted for the isochrone fitting.
As discussed in the next section, this mild systematic difference on age has, however, only a very minor effect on measuring the precession rate of the Galactic warp.     

\subsection*{Robustness of the warp-model fits and the systematic error of the precession rate.}
\par
The robustness of the warp-model fits on determinations of the LON for different age populations is discussed in Main, which considers several different effects.
 The first is the different choices of starting radius $R_{\rm s}$.
 Supplementary Figure\,5 shows the $R-Z$ distribution of sample stars with azimuth angle $10^{\circ}<|\phi|<90^{\circ}$. The warp signal is very clear in the two directions (warp down at negative azimuth and warp up at positive azimuth).
 The sample stars are divided into different radial bins  with a width of 0.5\,kpc.
 The median vertical distance, calculated at each radial bin, is largely close to the disk plane, i.e., $Z = 0$, for $R$ within 7.5\,kpc and tends to deviate from the disk plane beyond this radius.
 The starting radius is then set to 7.5\,kpc in our warp model (Equation\,1).
 We repeated the analysis of warp-model fits by choosing different values of $R_{\rm s}$ from 7 to 9\,kpc.
 The tests exhibited only a minor effect on determinations of the LON for different age populations.
 The overall change in the measurement of the trend between the LON and median age of different populations, that is , the precession rate $\omega$, was no more than 0.2 $\rm{km}\  s^{-1} \ \rm{kpc}^{-1}$.
 \par
 To test the effects arising from errors in the distance estimates of the Cepheids, the analysis in the Main was checked by changing the distances assuming a systematic error of 10\%, seven times larger than that examined by Cepheid--OC pairs.
The trend of the varying LON with median age for different populations held very well, and the overall effect on the precession-rate measurement was no more than 0.3\,$\rm{km}\  s^{-1} \ \rm{kpc}^{-1}$.
For the effects from age determinations, we first performed a similar test by changing ages and adopting the systematic error found by the check on OC--Cepheid pairs. 
The measured precession rate closely matches the value reported in the Main, roughly falling within the $1\sigma$ uncertainty.
Furthermore, two different priors were adopted to assign metallicities for the sample Cepheids when calculating their ages.
The first prior was based on a metallicity distribution function constructed from a small sample of Cepheids with [Fe/H] measured from high-quality high-resolution spectroscopy\citep{DS23}.
The second one simply took the mean value of this distribution ($\rm{[Fe/H]}=-0.07$) for all the Cepheids.
Under both priors, the measured warp-precession rates change by no more than 0.3\,$\rm{km}\  s^{-1} \ \rm{kpc}^{-1}$.
\par
We also checked for potential selection effects of the sample stars on the measurement of the precession rate.
The distant sample stars were more important than the nearby ones for constraining the model parameters, especially the LON, although the number of the former was much smaller than that of the latter.
To show that all the data points, especially the distant sample stars, were properly fitted by our 
best-fit models, detailed comparisons between observations and models as a function of different bins of azimuth range are shown in Supplementary Figure\,6, using the young (20-120\,Myr) and old (120-220\,Myr) populations as examples. 
Generally, the model predictions reproduce the observations at all distances very well.
We also repeated the fitting process by adding weights to the distant sample stars. All the results are quite similar to those found in the canonical case.
\par
Finally, we checked the effect of different quality cuts, better than $5$\% to $10$\%, on the distances used for the warp-model fitting analysis.
All the above tests result in no more than $0.2\ \rm{km}\  s^{-1} \ \rm{kpc}^{-1}$ changes in the measured warp-precession rate.
\par
To conclude, the above comprehensive checks demonstrate the robustness of our measurements of the precession rate of the Galactic warp and the overall potential systematic error is within 0.6\,km\,s$^{-1}$\,kpc$^{-1}$.

\subsection*{Constraining the disk-warp precession rate from the kinematic method.}
\par
Here we attempt to constrain the precession rate of the disk warp based on the canonical kinematic method from our Cepheid sample stars with high-quality radial velocity measurements from Gaia DR3.
For a kinematic-warp model with precession (see the details in Method of  ref\citep{Poggio20}), the vertical velocity distribution can be expressed as:
\begin{equation}
    V_{Z}=\left(\Omega_R -\omega\right)c(R-R_{\rm s})^{\alpha}cos(\phi-\phi_{w})\text{,}
\end{equation}
where
\begin{equation}
    \phi_w (t) = \phi_{0, w} + \omega (t-t_0)\text{.}
\end{equation}
Here, $\Omega_R$ represents the circular frequency of the adopted tracers at $R$, which can be calculated from the mean azimuthal velocity $\overline{V_{\phi}}$ over the radius $R$.
We set $\overline{V_{\phi}} = V_{\rm c}$ for these young Cepheids, given their negligible asymmetric drifts, due to their young and kinematically cold nature. 
The $V_{\rm c}$ parameter was adopted from a recent determination\citep{Zhou23} with a weak decline with increasing $R$: $V_{\rm c} (R)=234.04-1.83(R-R_{0})$ km\,s$^{-1}$.

\par
To determine $\omega$ kinematically, we first selected sample stars with reliable mean radial velocity measurements in Gaia DR3. 
Given the pulsational nature of Cepheids, the number of radial velocity measurements is required to be greater than eight to minimize the effects from pulsation.
Second, we focused on the sample stars with $7.5 < R < 16$\,kpc to ensure that we had notable warp signals. 
Here, $7.5$\,kpc corresponds to the starting radius of the geometric warp.
The distance cut was to ensure accurate vertical velocity, which is the key to measuring $\omega$ kinematically.
In total, 1,268 Cepheid stars were left after the two cuts.
The three-dimensional velocities ($V_{R}$, $V_{\phi}$ $V_{Z}$) in Galactocentric cylindrical coordinates are derived for these stars from their observed positions, proper motions, and radial velocities.

According to Equation\,4, $V_Z$ is expected to be a function of $\phi$ and $R$.
We first divided the sample stars into different azimuthal bins.
We focused on the azimuth range of $-70^{\circ}<\phi<50^{\circ}$ for which the number of stars is sufficiently large.
By choosing a width of $10^{\circ}$, we had 12 azimuthal bins in total.
For each azimuthal bin, the sample stars are further sub-divided into seven radial bins, with the first bin covering $R$ between 7.5 and 9\,kpc, and the remaining 6 radial bins having equal numbers of stars.
The median radius, azimuthal angle, and vertical velocity are calculated for 84 individual bins (Supplementary Figure\,7).
The kinematic model described by Equation\,4 was then fitted to these bins by adding an average source vertical velocity offset $V_{Z}^{\rm s}$.
The offset velocity adopted here was to correct for the possible non-zero vertical velocity at the starting radius.
Inspection of the full Cepheid sample across the entire outer disk led to an estimate of $V_{Z}^{\rm s}=-4.2\pm0.8$\,km\,s$^{-1}$.
In the end, the best fit yielded a warp-precession rate $\omega=-1.1\pm1.9$\,km\,s$^{-1}$\,kpc$^{-1}$.
The measured value of $\omega$ is consistent with our motion-picture measurement within the uncertainty.
However, due to the large measurement errors, we cannot rule out the possibility that the disk warp is not precessing, based solely on this analysis. 

The original data points and the bin results are shown in Supplementary Figure\,7. 
In general, the scatter of the vertical velocities is quite large, as high as 5 to 10 km\,s$^{-1}$, for all radial bins and azimuthal directions.
Moreover, the distribution of the vertical velocity is not smooth but has notable jumps or dips, for example the mean vertical velocity at $R \sim 13$\,kpc of $-10^{\circ} < \phi < 0^{\circ}$.
The above behaviours, as well as the non-zero $V_{Z}^{\rm s}$ were at least partly caused by another vertical disequilibrium source (other than the warp) -- the well-known snail-shell or phase spiral\citep{Antoja18}.
For these reasons, the kinematic method cannot precisely determine the precession rate with a high accuracy (better than a few \,km\,s$^{-1}$). 
However, the overall trends of the vertical velocities clearly disfavor the large precession rate in the prograde direction found by recent estimates of $10.83 \pm 0.03\ {\rm (statistical)} \pm 3.20\ {\rm (systematic)}$\,km\,s$^{-1}$\,kpc$^{-1}$ from Gaia giant stars\citep{Poggio20}, and of 5.9-11.4\,km\,s$^{-1}$\,kpc$^{-1}$ at the guiding center radius 10-14\,kpc using Cepheid stars\citep{Dehnen23} similar to those used here.

\subsection*{Constraining $q_{\Phi}$ from the disk-warp precession rate.}
Following ref.\citep{SS06}, the Galactic disk was assumed to be composed of a series of rigid rings (or annuli).
The precession rate of each ring at radius $R$ and inclination angle $i$ (relative to the symmetry plane of the torquing source) was calculated analytically from the torques provided by the massive disk and DM halo. Details are described in Appendix\,A of ref.\citep{SS06}.
In short, the precession rate can be calculated by:
\begin{equation}
\omega = \frac{\langle T \rangle}{L\sin{i}} = \frac{\langle T \rangle}{RV_{\rm c}\sin{i}},
\end{equation}
where $V_{\rm c}$ is the rotation curve, again adopted from ref.\citep{Zhou23}, and  $\langle T \rangle$ is the azimuthally averaged torque on the rigid ring:
\begin{equation}
\langle T \rangle = \frac{1}{2} T_{\rm max} = \frac{1}{2} rF_{\theta} = -\frac{1}{2}\frac{\partial \Phi}{\partial \theta},
\end{equation}
where $\theta = \pi/2 - i$. 
The precession rates computed by this simple rigid-ring model are well-validated by $N$-body numerical simulations which properly consider the self-gravity and random motions of disk stars\citep{SS06}$^{,}$\citep{Jeon09}.
Note that $\cos i \approx 1$ is assumed in the following derivations of $\omega_{\rm disk}$ and $\omega_{\rm halo}$, given the small value of $i$ (only about $2.9^{^\circ}$ at $R \sim 15$\,kpc). 
\par{}
The massive disk was assumed to contain two exponential components: a thin disk and a thick disk.
By substituting their potentials (see Equation\,7 of ref.\cite{SS06}) into Equations 6 and 7, the precession rates contributed by the two disks at large radius can be calculated by:
\begin{equation}
\begin{aligned}
    \omega_{\rm disk} \approx \ -\frac{3}{2}\frac{3R_{\rm d}^2}{R^2}\frac{GM_{\rm d}}{R}(RV_{\rm c})^{-1}\text{,}\\
\end{aligned}
\end{equation}
where $R_{\rm d}$ and $M_{\rm d}$ represent the disk scale length and the total disk mass. The total disk mass can be calculated by $M_{\rm d} = 2\pi\Sigma_0R_{\rm d}^2$, where $\Sigma_0$ represents the disk central surface density.
For the two disks, most of their parameters have been well measured. 
For the surface densities, we adopted the local measurements of $\Sigma_{R_{0}, \rm{thin}}=30.4\ M_{\odot}\ \rm{pc}^{-2}$ (refs.\citep{Flynn06}$^{, }$\citep{Bovy13}) and $\Sigma_{R_{0}, \rm{thick}}=7\ M_{\odot}\ \rm{pc}^{-2}$ (ref.\citep{Flynn06}) for the thin disk and thick disk, respectively. 
In principle, the scale length of both the thin and thick disks can be derived by fitting a mass model to the Galactic rotation curve\citep{Zhou23}. However, they are strongly degenerate with the parameters of the DM halo. To break this degeneracy, the scale length of the thin disk and the total disk mass were fixed, based on independent measurements, when calculating the disk-warp precession contribution from the Galactic disks. The current measurements of disk scale lengths for both disks have not yet reached convergence. Therefore, we set the scale length of the thin disk in the range of 2 to 4\,kpc, an interval that covers almost the full range of existing measurements\citep{Bland16}. 
The scale length of the thick disk was then calculated to ensure that the total disk mass matches the recent direct dynamical measurement\citep{Bovy13}: $M_{\rm d}^{\rm thin + thick} = 5.1\times10^{10}\ M_{\odot}$.
In this way, the contributions of the two disks on the warp precession were computed under different choices of $R_{d, {\rm thin}}$, reaching about $-0.98$\ \,km\,s\,$^{-1}$\,kpc$^{-1}$ ($R_{d, {\rm thin}} = 4$\,kpc; see Figure 2a) to  $-0.55$\ \,km\,s\,$^{-1}$\,kpc$^{-1}$ ($R_{d, {\rm thin}} = 2$\,kpc; see Supplementary Figure\,8) at $R = 14$\,kpc.
After subtracting the disk contributions from the measured precession rates, the residuals exhibited negative values spanning from  $-1.5$ to $-1.0$\ \,km\,s\,$^{-1}$\,kpc$^{-1}$ (Extended Data Figure\,3), suggesting that there is a substantial contribution from an oblate DM halo.

To quantitatively constrain the oblateness, an Navarro-Frenk-White model\citep{NFW96} (modified by adding a flattening parameter\citep{K15}), rather than the torus model in ref.\citep{SS06}, was adopted to represent the density profile of the DM halo.
The potential of this model can be  expressed as:
\begin{equation}
    \Phi=-\frac{4\pi G \rho_{s} r_{s}^3}{\sqrt{R^2+\frac{z^2}{q_{\Phi}^2}}}\ln(1+\frac{\sqrt{R^2+\frac{z^2}{q_{\Phi}^2}}}{r_{s}}),
\end{equation}
\noindent 
where $r_{s}$ and $\rho_{s}$ represent the scale radius and the characteristic DM density, respectively, and $q_{\Phi}$ is the flattening parameter (minor-to-major axis ratio) that describes the shape of the DM halo. 
By again substituting this potential into Equations 6 and 7, the precession due to the DM halo can be calculated by:
\begin{equation}
    \omega_{\rm halo} \approx \ -\frac{1}{2}(\frac{1}{q_{\Phi}^2}-1) 4\pi G \rho_{s}\ r_{s}^3\ \frac{1}{V_{\rm c}}\frac{1}{R^2}(\ln(1+\frac{R}{r_{s}})-\frac{R}{R+r_{s}}).
\end{equation}
 When calculating the precession rates from DM halo, the scale radius $r_{s}$ and the characteristic density $\rho_{s}$ are required to be known.
Rather than adopting fixed values from literature, the two parameters were determined by fitting a mass model of the Milky Way to the latest measurements of the Galactic rotation curve  from ref.\citep{Zhou23}.
This mass model consists of three components: two disks, a DM halo, and a bulge.
The disks and DM halo are the same to these adopted for computing the warp-precession rates.
For the bulge, a Plummer bulge ($\rho_{\rm bulge} = \frac{3b^2M_{\rm b}}{4\pi(r^2+b^2)^{5/2}}$) is adopted with $b = 0.3$\,kpc and $M_{\rm b} = 1.067 \times 10^{10} M_{\odot}$ (ref.\citep{Zhou23}).
Note that this bulge does not contribute to the warp precession due to its spherical nature.
At each choice of $R_{d, {\rm thin}}$, the flattening parameter $q_{\Phi}$ was then derived by fitting the above warp-precession model (Equations 8 and 10) to the measured precession rates (see Figure 2 and Supplementary Figure 8 as examples) with DM density parameters $r_{s}$ and $\rho_{s}$ already known from fitting the mass model to Galactic rotation curve (see  Supplementary Figure 9 as an example).
All fits were performed by Markov chain Monte Carlo approach.
The 16th and 84th percentiles from the resulting posterior probability distribution function (PDF) were adopted to define the interval for $q_{\Phi}$.
By taking a step of $0.1$\,kpc at $R_{d, {\rm thin}}$, a series of intervals are found.
The final range for $q_{\Phi}$, determined as [0.84, 0.96], was given by the combined set of these obtained intervals.

\par
The current analysis focused on exploring the oblateness of the DM halo, as it primarily influences the warp precession\citep{Jeon09}.
we will continue to improve the entire analysis by 1) studying more complicated halo models (for example, a triaxial halo with two flattening parameters and an orientation angle as mentioned in a recent study\citep{HAN23}.) with advanced $N$-body numerical simulations, and 2) exploring alternative mechanisms that may contribute to the warp precession.

\par
Note that some authors in the literature constrain the shape of DM halo using the axis ratio of the isodensity contour 
$q_{\rho}$.
Here, the $1-q_{\Phi} \approx \frac{1}{3}(1-q_{\rho})$ transformation was adopted for those estimates using $q_{\rho}$ (Figure 2(b); ref.\citep{BT08}).

\begin{table}[H]
    \setlength{\abovecaptionskip}{0.1cm}
    \setlength{\belowcaptionskip}{0.2cm}
    \begin{center}
    \caption{\label{tab:l1}Measured LON in each age bin yielded by the best-fit model} 
    \begin{tabular}{cccc}
        \hline
        Age bin & Number & Median age & LON \\
        (Myr) &   & (Myr) & (degrees)\\
        \hline
        (0, 100) & 1175 & 79.6$\pm$19.9 & 5.59$\pm$1.71\\
        (20, 120) & 1603 & 89.5$\pm$23.3 & 6.14$\pm$1.34\\
        (40, 140) & 1806 & 98.6$\pm$24.5 & 7.56$\pm$1.11\\
        (60, 160) & 1800 & 106.3$\pm$24.8 & 10.26$\pm$1.02\\
        (80, 180) & 1557 & 113.9$\pm$24.1 & 10.31$\pm$1.02\\
        (100, 200) & 1123 & 128.1$\pm$23.8 & 11.45$\pm$1.14\\
        (120, 220) & 726 & 144.1$\pm$24.3 & 13.29$\pm$1.34\\
        (140, 240) & 465 & 161.4$\pm$26.5 & 14.93$\pm$1.70\\
        \hline
    \end{tabular}
    \end{center}
    Note: the uncertainties in the median age were calculated as the standard deviation of ages at each bin.
\end{table}

\begin{figure*}
    \begin{center}
        \includegraphics[scale=0.15, angle=0]{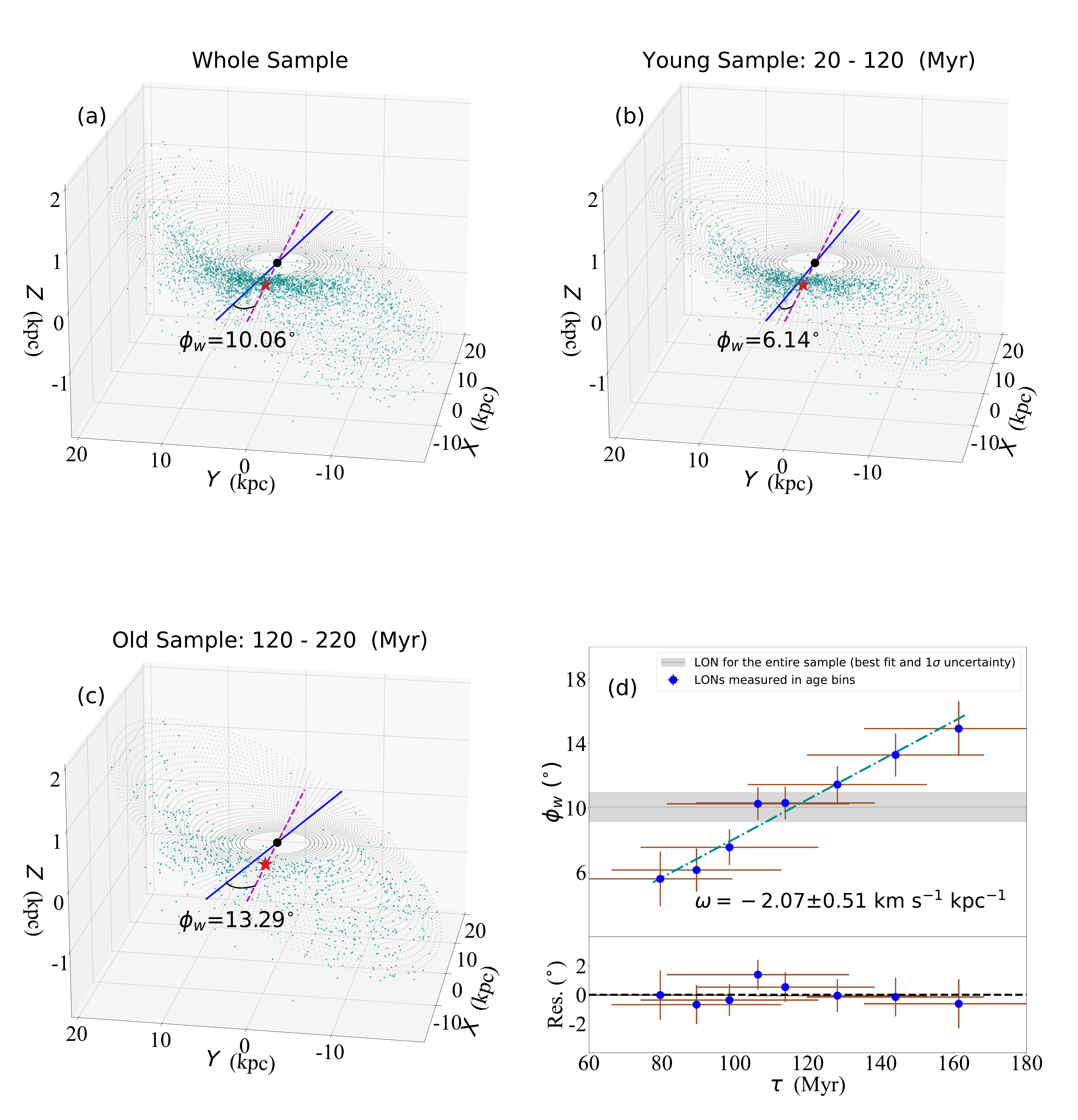}
        \caption{\textbf{The Milky Way's 3D disk warp and its precession traced by Cepheids.} 
        (a) The disk-warp structure revealed by our full sample of 2,613 Cepheids (cyan dots). The grey grid is the best-fitting model (described by Equation\,1). The blue line denotes the LON with the best-fit $\phi_w = 10.06\pm0.93^{\circ}$.
        The purple dashed line connects the Sun (red star) and Galactic center (black dot).
        (b) The disk-warp structure revealed by the young Cepheid sample (20 to 120\,Myr). 
        The best-fitting $\phi_w$ for the LON is $6.14\pm1.34^{\circ}$.
        (c) The disk-warp structure revealed by the old Cepheid sample (120 to 220\,Myr).
        The best-fitting $\phi_w$ for the LON is $13.29\pm1.34^{\circ}$.
        (d) Measured LON as a function of median age for different bins of Cepheids. The error bars represent $1\sigma$ confidence regions. The cyan dotted-dashed line represents a linear fit to the data points, yielding a precession rate: $\omega = -2.07 \pm 0.51$\,km\,s$^{-1}$\,kpc$^{-1}$.
        The residuals (Res.) between the measured LON and the linear fit are shown in the bottom part of the panel.}
    \end{center}
\end{figure*}

\begin{figure*}
    \centerline{\includegraphics[scale=0.425, angle=0]{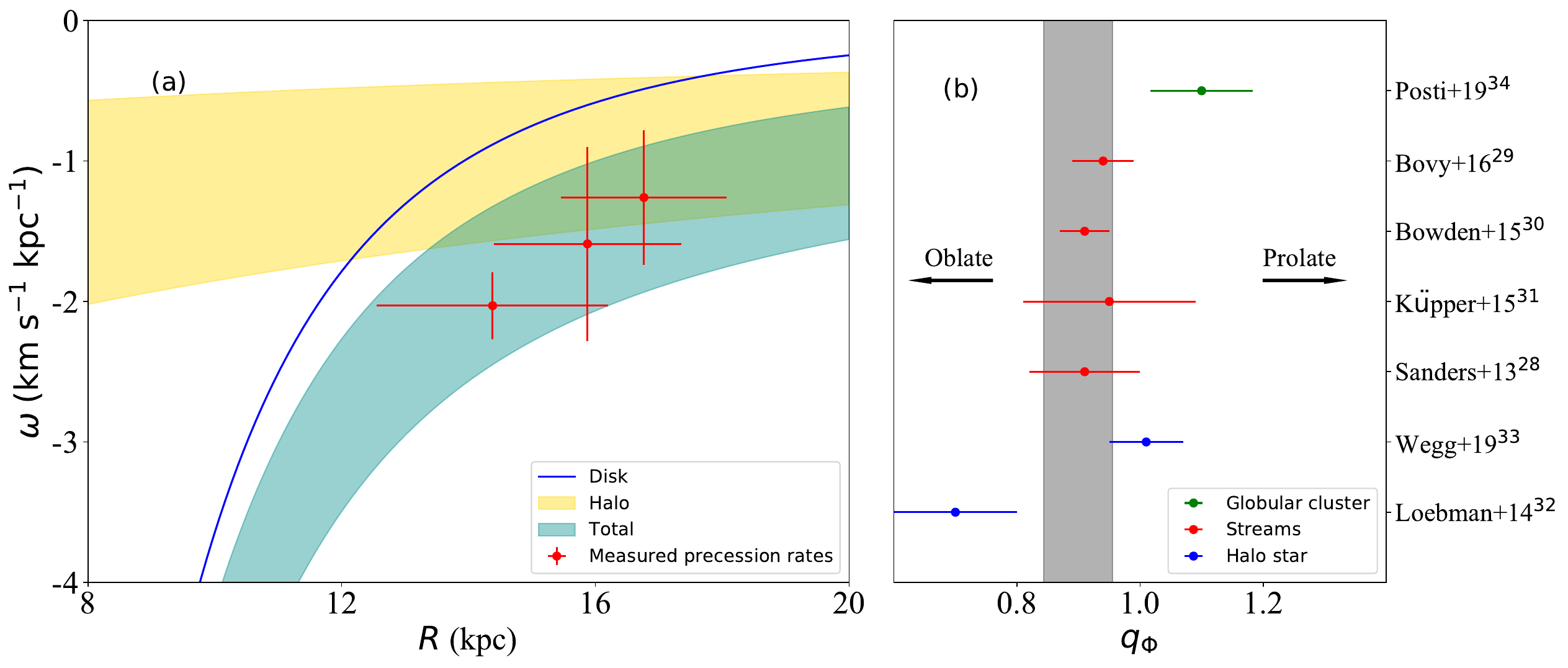}}
    \caption{\textbf{Constraining the shape of the DM halo from the measured precession rates.}  (a) The precession rate of the disk warp as a function of Galactocentric distance $R$. The error bars denote $1\sigma$ confidence regions. As an example, the blue line denotes the contributions from both the thin and thick disks on the warp's precession by adopting $R_{d, \rm{thin}}=4\ \rm{kpc}$. The dark cyan-shaded region represents the best-fit radially dependent precession rates of the disk-warp to the data points (red dots), which is the sum of the contributions from the Galactic disk (blue line) and the DM halo (gold shaded region). For other choices of $R_{d, \rm{thin}}$, the fitting results are presented in Supplementary Figure 8.
    (b) The best-fitting interval of the shape of the DM halo. The flattening (minor-to-major axis ratio) of the equipotential surface $q_{\Phi}$ is from 0.84 to 0.96 (grey shaded region). Recent estimates of $q_{\Phi}$ from the kinematics of globular clusters (green dots) stellar streams (red dots), and the 6D distributions of halo stars (blue dots), are overplotted for comparison. The error bars are the $1\sigma$ standard deviations.}
\end{figure*}

\section*{Data Availability}
The Cepheids data used in this paper are publicly available from Gaia archive: {\url{https://archives.esac.esa.int/gaia}}. The other data supporting the plots in this paper and other findings of this study are available from the corresponding authors upon reasonable request.

\section*{Code Availability}
We use standard data analysis tools in the Python environments, including methods in astropy, numpy, matplotlib, scipy and emcee. All these packages are publicly available through the Python Package Index ({\url{https://pypi.org}}).
Specifically, the fit analysis in this study is performed using the Python package scipy.curve\_fit and emcee.

\section*{Acknowledgements}
YH acknowledges the National Science Foundation of China (NSFC) with Nos. of 11903027 and 11833006 and the National Key RD Program of China No. 2019YFA0405503. 
HWZ acknowledges the National Key RD Program of China No. 2019YFA0405504, and
the NSFC with Nos.of 12090040, 12090044.
JFL acknowledges support from the NSFC through grant Nos. of 11988101 and 11933004, and support from the New Cornerstone Science Foundation through the New Cornerstone Investigator Program and the XPLORER PRIZE.
JS acknowledges support from NSFC through grant Nos. 12025302 and 11773052, support from the “111” Project of the Ministry of Education of China through grant No. B20019, and support from China Manned Space Project through grant No. CMS-CSST-2021-B03. 
JS also acknowledges support from the Gravity Supercomputer at the Department of Astronomy, Shanghai Jiao Tong University, and the Center for High Performance Computing at Shanghai Astronomical Observatory. 
TCB acknowledges partial support for this work from grant PHY 14-30152; Physics Frontier Center/JINA Center for the Evolution of the Elements (JINA-CEE), awarded by the US National Science Foundation, and from OISE-1927130: The International Research Network for Nuclear Astrophysics (IReNA),
awarded by the US National Science Foundation. 
We also express thanks for the valuable suggestions and comments from the ``Frontier discussion of top sciences'' regularly held at the National Astronomical Observatories, Chinese Academy of Sciences, where we presented the main results of this study on October 28, 2022.  
This work presents results from the European Space Agency (ESA) space mission Gaia. Gaia data are being processed by the Gaia Data Processing and Analysis Consortium (DPAC). Funding for the DPAC is provided by national institutions, in particular the institutions participating in the Gaia MultiLateral Agreement (MLA). The Gaia mission website is {\url{https://www.cosmos.esa.int/gaia}}. The Gaia archive website is {\url{https://archives.esac.esa.int/gaia}}.

\section*{Contributions}
YH contributed to the design of this project and writing of the final paper;
QKF contributed to the sample preparation, modelling and data analysis, and wrote the manuscript together with YH;
TK contributed to the data analysis and revisions of the text;
HWZ contributed to the project planning and research support;
JFL contributed to the design of this project and revised the text;
TCB contributed to the interpretation and revisions of the text;
JS contributed to the theoretical computation of the warp precession rate, interpretation of the result, and text revision;
YJL contributed to the interpretation of the result;
SW contributed to the data analysis and revisions of the text;
HBY contributed to the data analysis and revisions of the text.

\section*{Competing interests}
The authors declare no competing interests.

\section*{Extended Data}

\begin{figure*}
    \begin{center}
    \includegraphics[scale=0.8, angle=0]{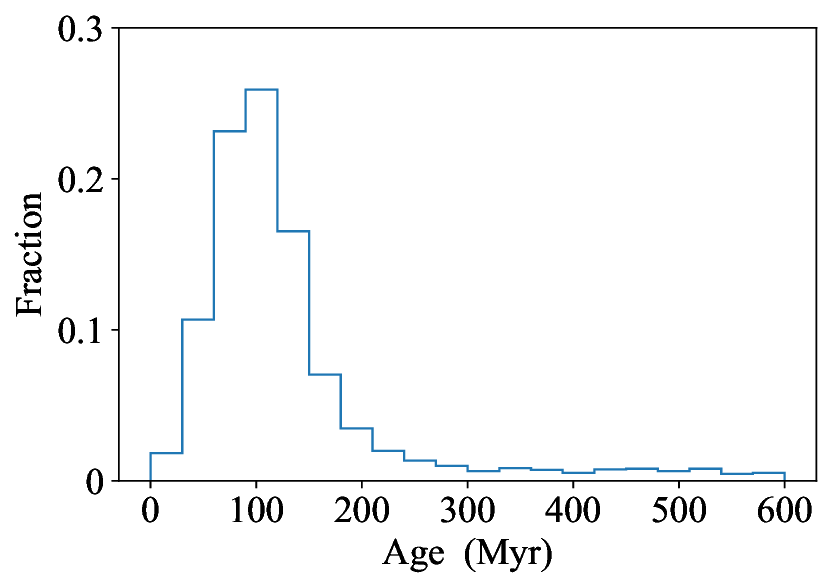}\\
    \end{center}
    \small{\textbf{Extended Data Figure 1. The age distribution of our final Cepheid sample.} Their ages are derived by the PAZ relation. Most of our sample stars are younger than 200 Myr.}
\end{figure*}

\begin{figure*}
    \begin{center}
    \includegraphics[scale=0.35, angle=0]{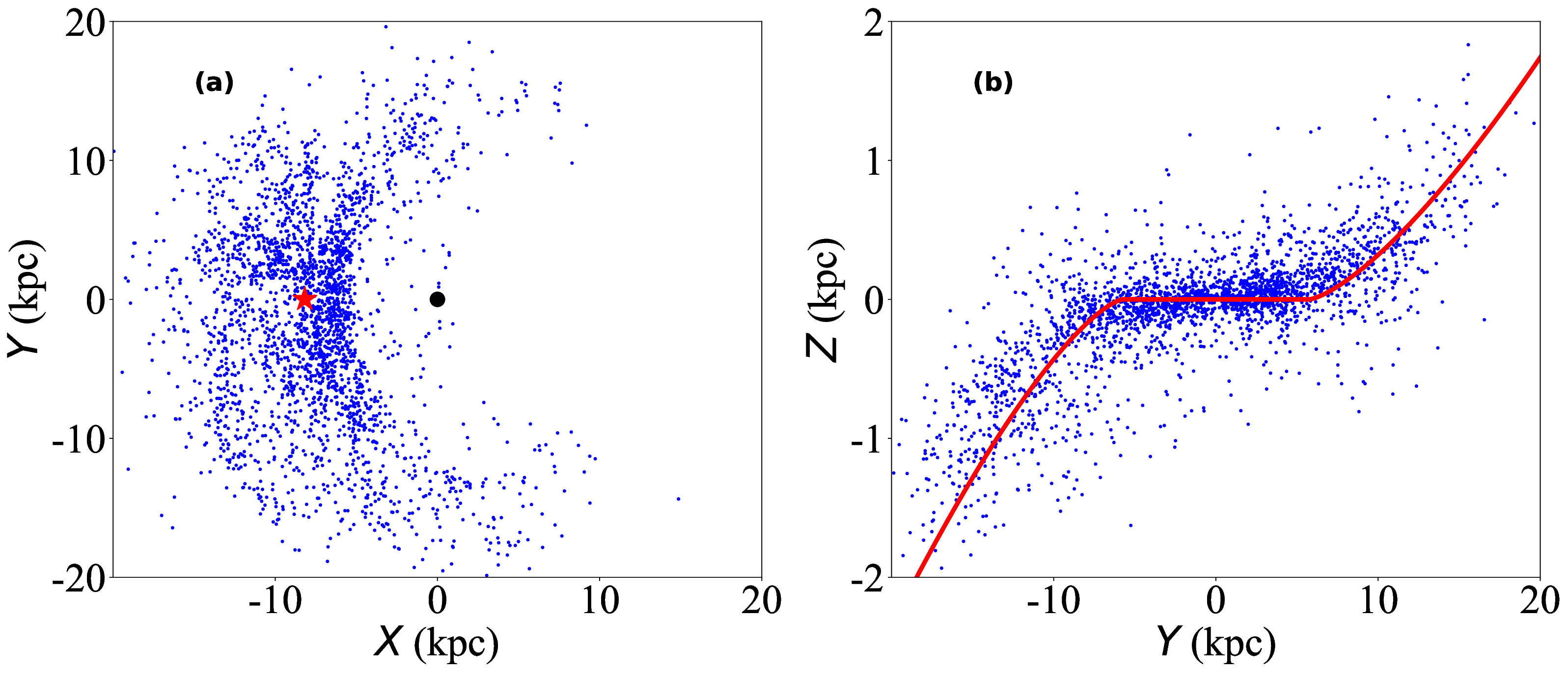}\\
    \end{center}
    \small{\textbf{Extended Data Figure 2. The spatial distribution of the final sample of 2,613 Cepheids.} (a) The $X-Y$ projection. The black dot and red star represent the location of the Galactic centre and the Sun, respectively. (b) The $Y-Z$ projection. The red line denotes the best-fit model with Galactic azimuth angle $\phi=\pm50^{\circ}$. Note that the warp amplitude is exaggerated, as the $Y-Z$ axes are not on the same scale.}
\end{figure*}

\begin{figure*}
    \begin{center}
    \includegraphics[scale=0.6, angle=0]{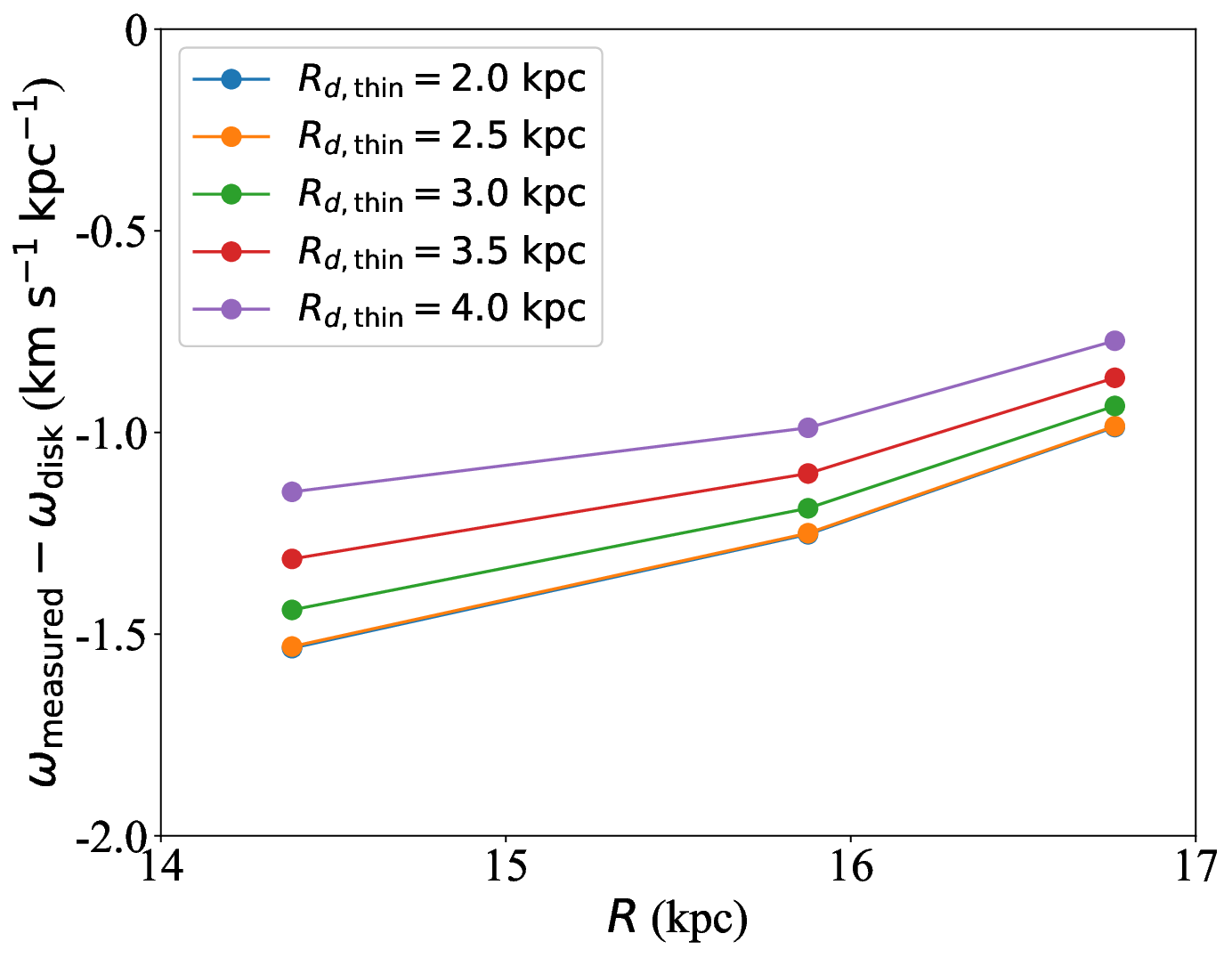}\\
    \end{center}
    \small{\textbf{Extended Data Figure 3. The residual precession rates, after subtracting the disk contributions.} In the range of 2 to 4 kpc, all the residual prcession rates are clearly non-zero.}
\end{figure*}

\section*{Supplementary information}

\subsection*{Contents}

Supplementary figures \textbf{1}-\textbf{10}

\counterwithin{figure}{section}

\begin{figure*}
    \begin{center}
    \includegraphics[scale=1, angle=0]{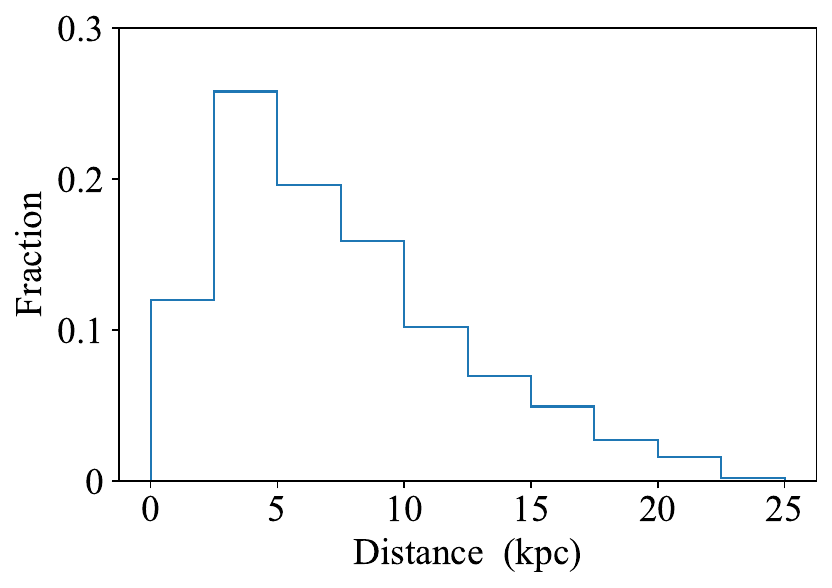}\\
    \end{center}
    \small{\textbf{Supplementary Figure 1}: The distance distribution of our final Cepheid sample. Their distances are derived from PW relation. Most of the sample stars are located within 20 kpc.}
\end{figure*}

\begin{figure*}
    \begin{center}
	\includegraphics[width=8cm]{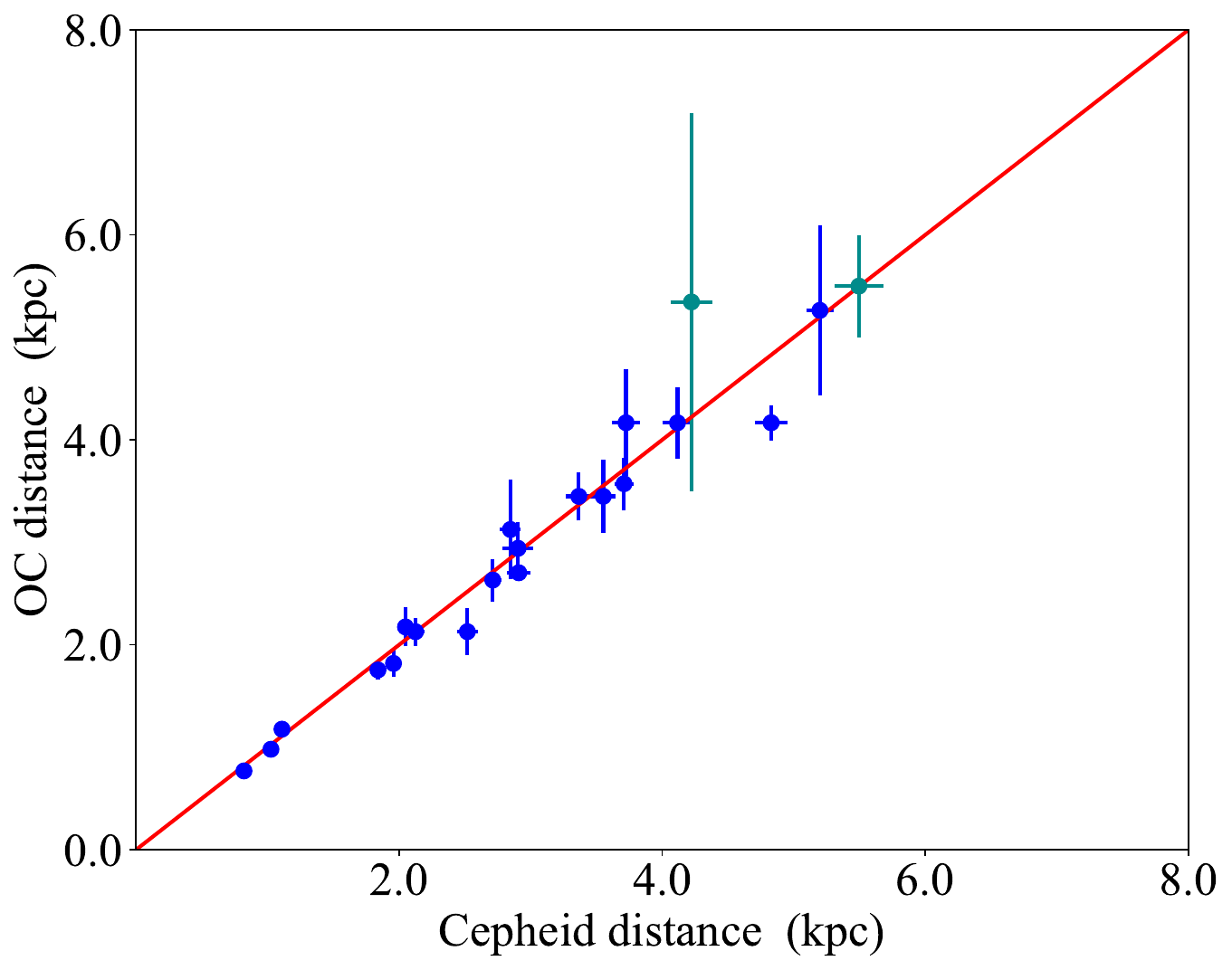}
	\includegraphics[width=8cm]{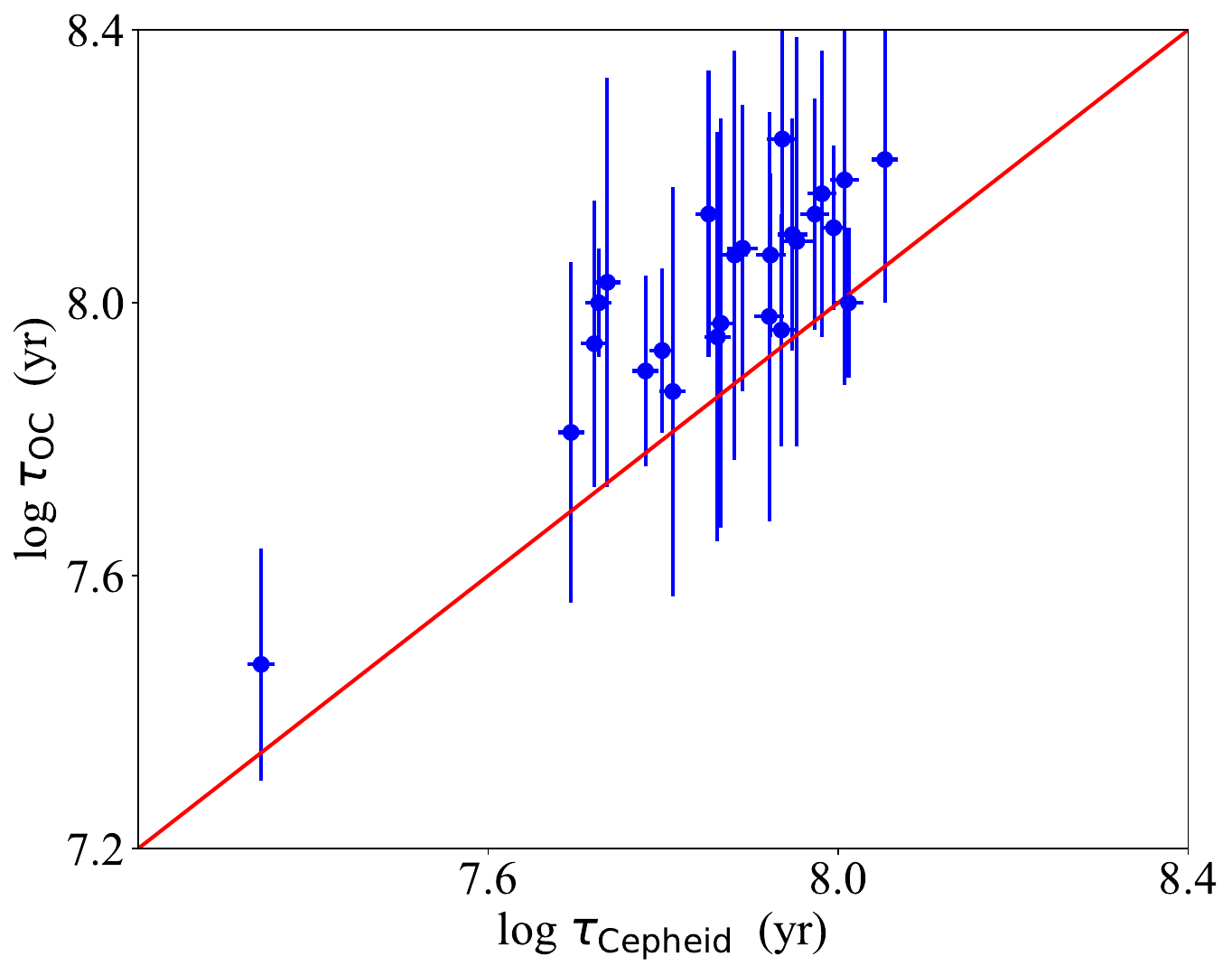}\\
	\end{center}
    \small{\textbf{Supplementary Figure 2}: The validations of distances and ages of our Cepheids sample using OCs. Left panel: Comparison between Cepheid distances from the PW relations and the OC's distances from the mean parallax measurements of OC member stars (blue dots) and the isochrone fitting (dark cyan dots). Right panel: Comparison between Cepheid ages from the PAZ relation and the OC ages from isochrone fitting, all on a logarithmic scale. In both panels, the error bars represent $1\sigma$ uncertainties.}
\end{figure*}

\begin{figure*}
    \begin{center}
    \includegraphics[scale=1, angle=0]{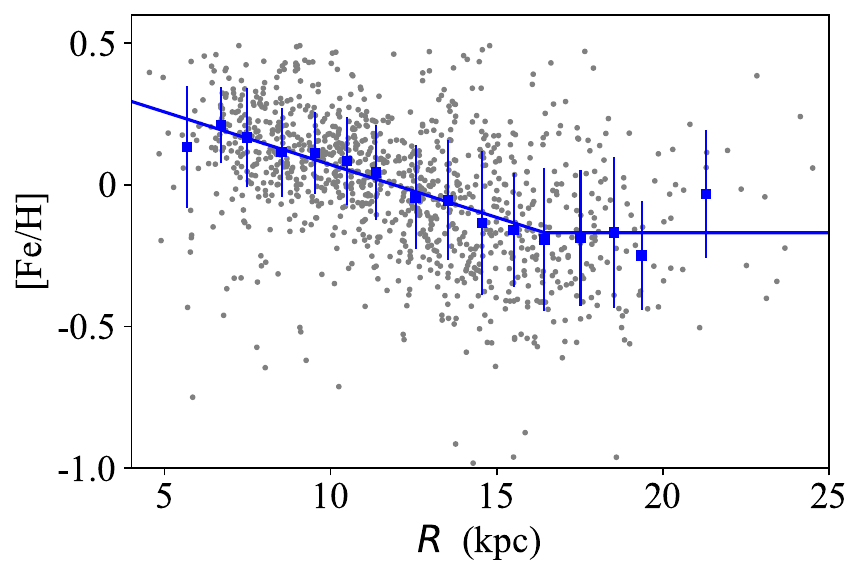}\\
    \end{center}
    \small{\textbf{Supplementary Figure 3}: Radial metallicity distribution of Cepheid sample stars with metallicity measured from the Gaia light curves. The blue squares and error bars represent the median values and $1\sigma$ standard deviations of [Fe/H] in individual radial bins, after $2\sigma$ to $3\sigma$ clipping.
    The blue lines indicate the best-fit models to the radial median metallicity, as described by Equation\,3.}
\end{figure*}

\begin{figure*}
    \begin{center}
    \includegraphics[scale=1, angle=0]{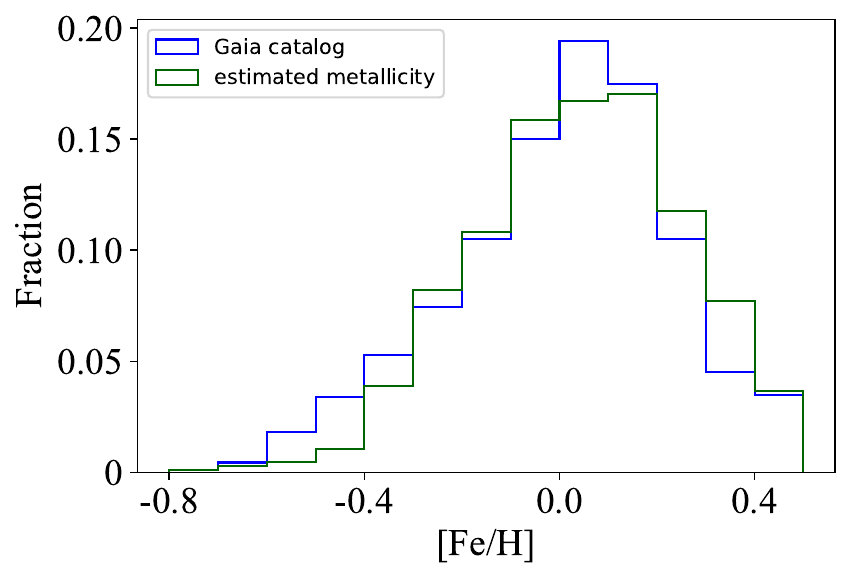}\\
    \end{center}
    \small{\textbf{Supplementary Figure 4}: The metallicity distribution of our Cepheid sample. 
    The blue line denotes the distribution of the one-third of sample stars with [Fe/H] values from the Gaia catalog, while the dark green line denotes the distribution of the remaining two-thirds of sample stars with [Fe/H] estimated from the radial metallicity distribution (see Methods).}
\end{figure*}

\begin{figure*}
    \begin{center}
    \includegraphics[scale=0.7, angle=0]{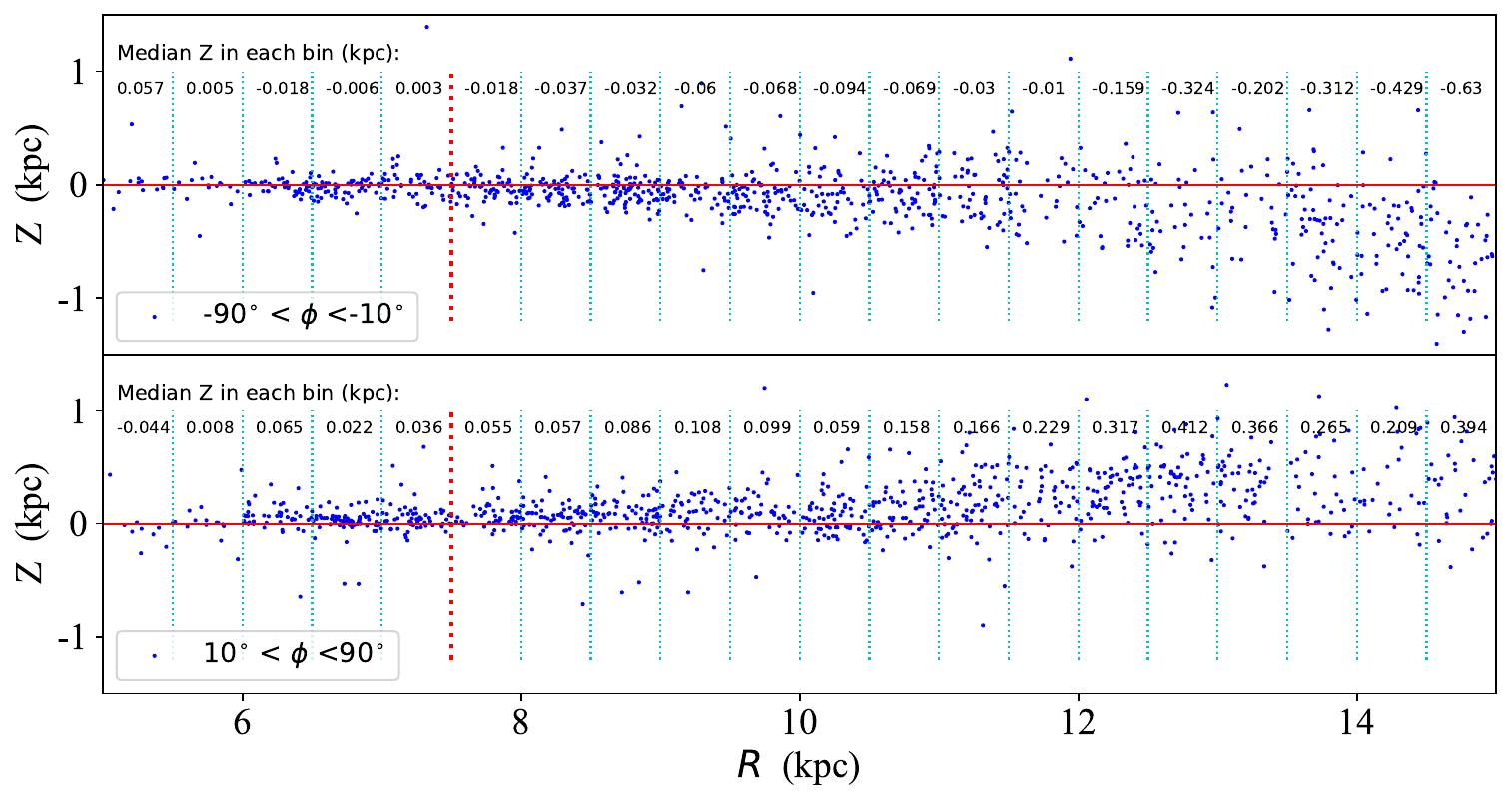}\\
    \end{center}
    \small{\textbf{Supplementary Figure 5}: The $R-Z$ distribution of the sample stars with $-90^{\circ}<\phi<-10^{\circ}$ (southern warp; upper panel) and with $10^{\circ}<\phi<90^{\circ}$ (northern warp; lower panel). The number marked in the top part of each subplot denotes the median $Z$ of each radial bin divided by the dashed cyan lines. The vertical thick dashed red line indicates the value of starting radius $R_{\rm s} = 7.5$\,kpc adopted in our warp model.}
\end{figure*}

\begin{figure*}
    \begin{center}
        \includegraphics[scale=0.5, angle=0]{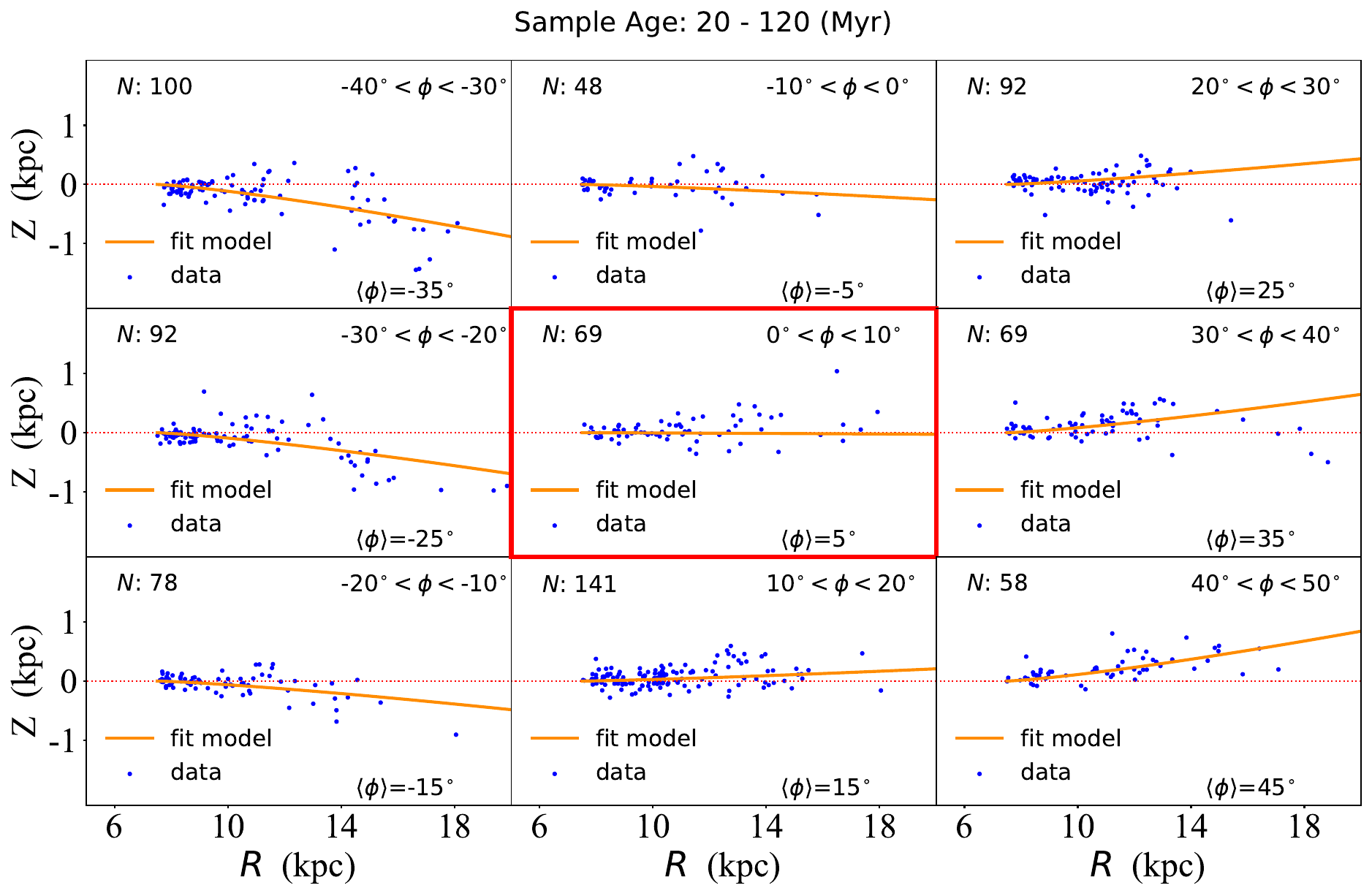}
        \includegraphics[scale=0.5, angle=0]{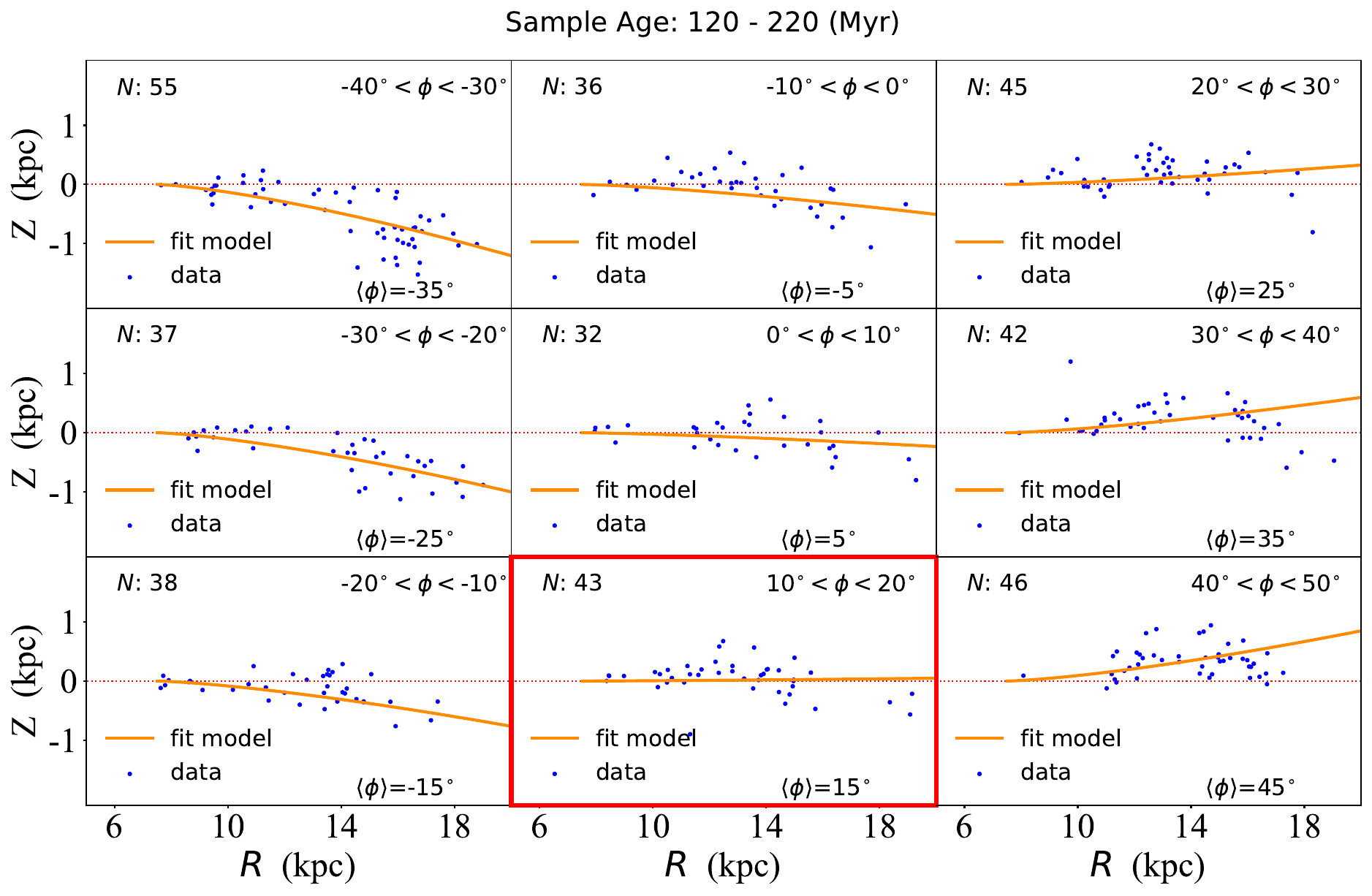}\\
    \end{center}
    \small{\textbf{Supplementary Figure 6}: Comparison between the observations (blue data points) of the $R - Z$ trend starting at $R = 7.5$\,kpc and the one (orange lines) predicted by best-fit models for different ranges of azimuth angle for a young population (20-120\,Myr; upper panel) and an old population (120-220\,Myr; lower panel) as examples.
    The range of azimuth and its median in each subplot are marked in the top-right and bottom-right corners, respectively. 
    The number of data points is labelled in the top-left corner. 
    Generally, all the model-predicted trends agree very well with that of observations for all sub-plots in both the upper and lower panels.
    The red boxes highlight the azimuth bins without warp signals that indicate the LON $\phi_w$ is within the azimuth range of this bin.
    Clearly, the LON of the young population (20-120\,Myr) is about $\phi_w \sim 5^{\circ}$, much smaller than $\phi_w \sim 15^{\circ}$ found for the old population (120-220\,Myr).}
\end{figure*}

\begin{figure*}
    \begin{center}
        \includegraphics[scale=0.54, angle=0]{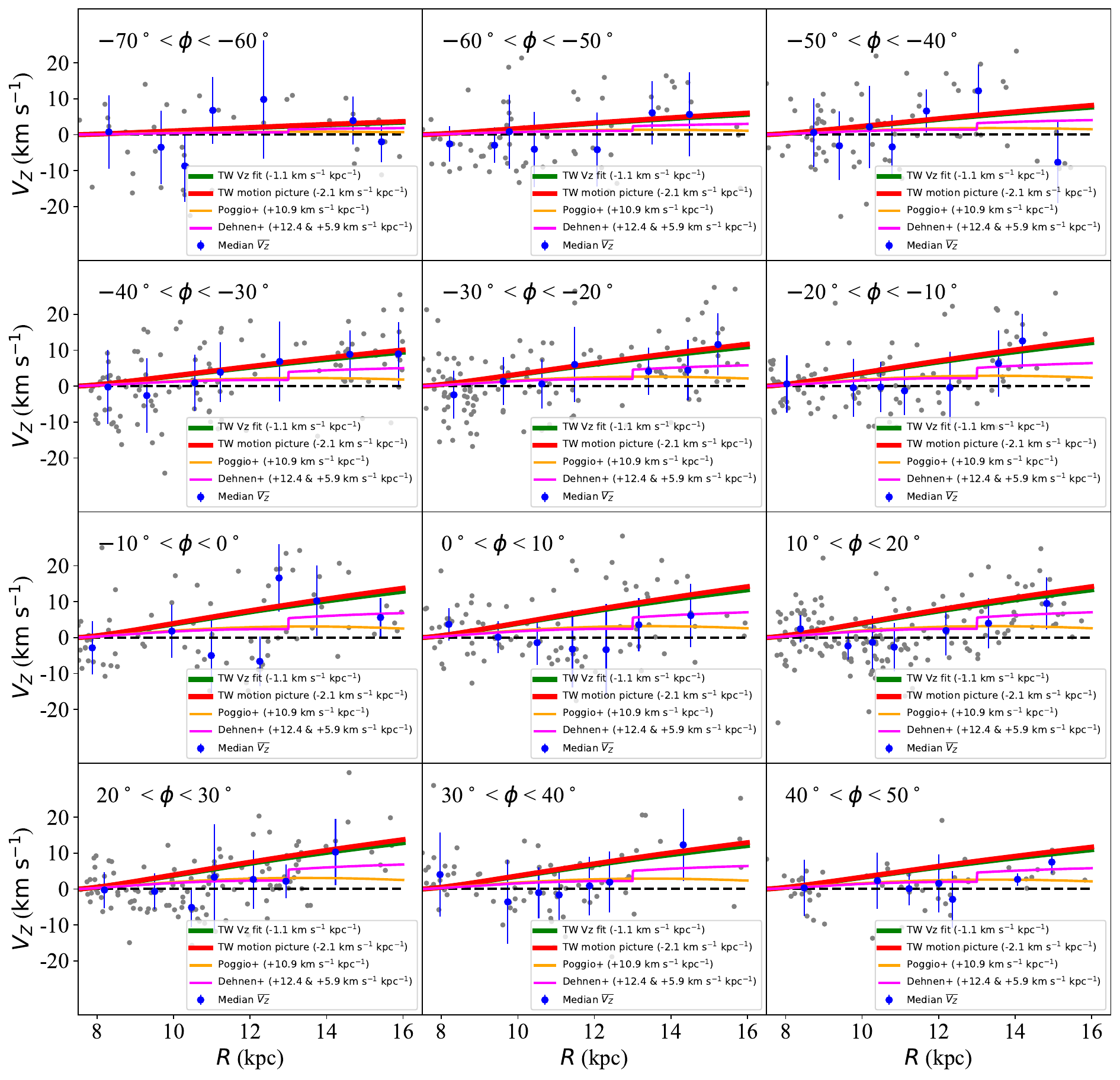}
    \end{center}
    \small{\textbf{Supplementary Figure 7}: Comparisons between the observations of the radial trend of $V_Z$ starting at $R = 7.5$\,kpc and the one (green lines) predicted by best-fit models (i.e., $\omega = -1.1$\,km\,s$^{-1}$\,kpc$^{-1}$) for different ranges of azimuthal directions. The radial trend of $V_{Z}$ predicted by other measured values (listed in legend) of disk-warp precessions are also shown. The blue dots and error bars denote median values and $1\sigma$ standard deviations of vertical velocities in each radial bin. The parameters of the geometric warp are set to those found in this study.
    The grey points represent the original data points used by the radial binning.
    The range of azimuth is marked in the top-left corner of each subplot.
    Note that an offset of $-V_{Z}^{\rm s} = 4.16$\,km\,s$^{-1}$ is added on the vertical velocity of the stellar tracers.}
\end{figure*}

\begin{figure*}
    \begin{center}
        \includegraphics[scale=0.32, angle=0]{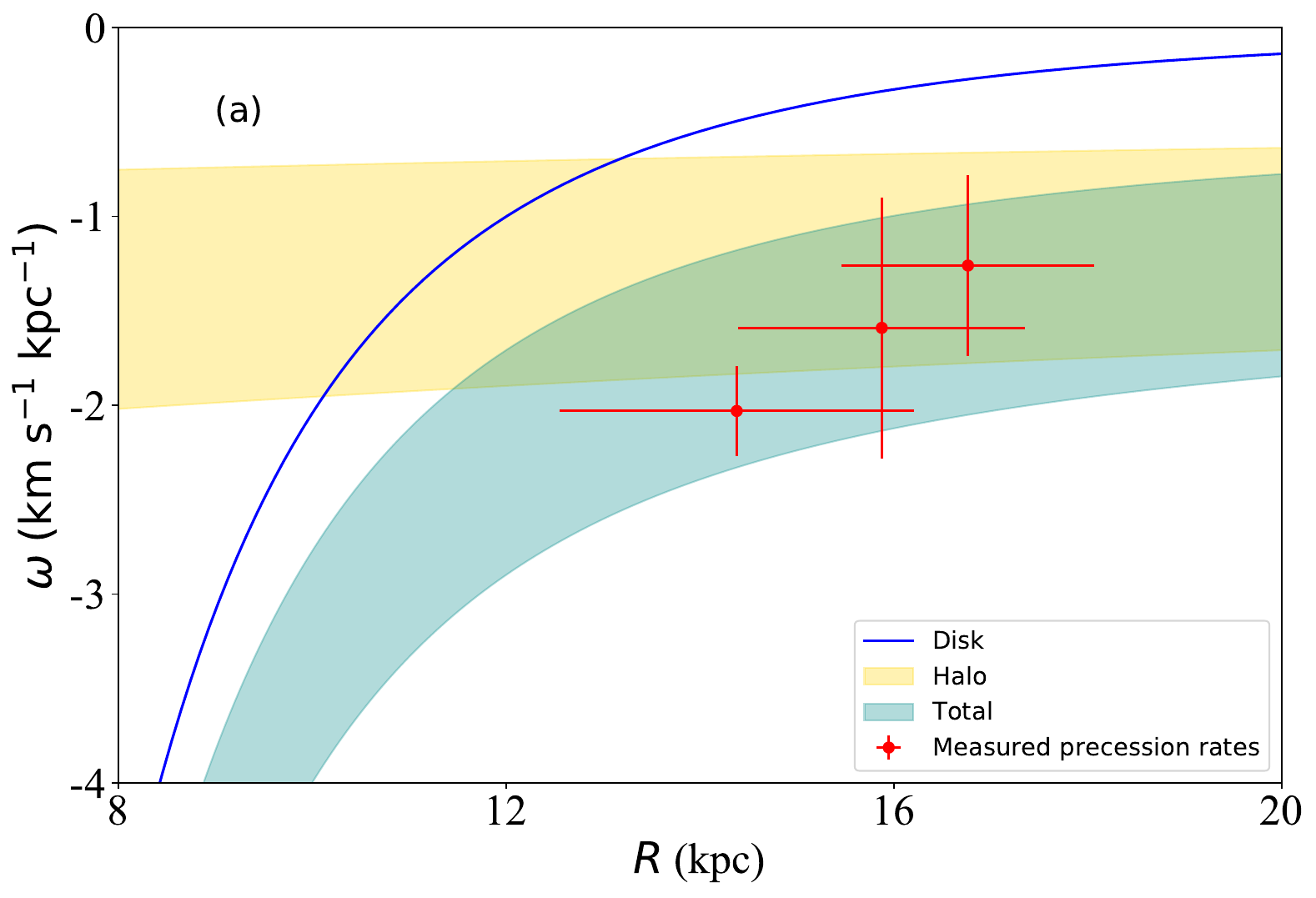}
        \includegraphics[scale=0.32, angle=0]{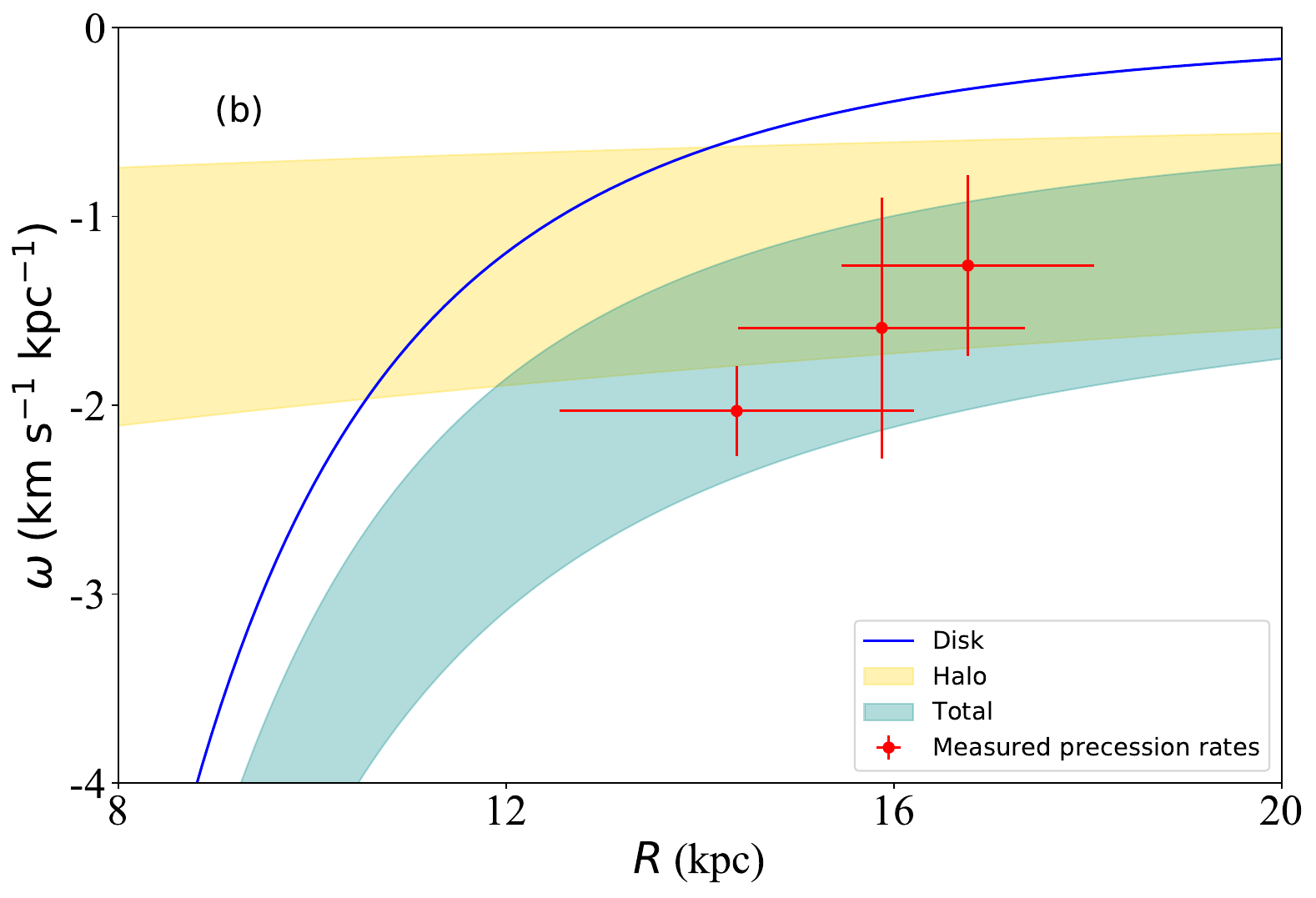}
    \end{center}
    \small{\textbf{Supplementary Figure 8}: Constraining the shape of the DM halo from the measured precession rates under alternative values of thin disk scale length. Similar to Figure 2(a) but for $R_{d, {\rm thin}} = 2$\,kpc (left panel) and $R_{d, {\rm thin}} = 3$\,kpc (right panel).
    The blue line, golden shaded regions, and dark cyan-shaded regions represent the contributions on disk-warp precession rates from the disk, the DM halo, and their sum, respectively. All error bars denote $1\sigma$ confidence regions.}
\end{figure*}

\begin{figure*}
    \begin{center}
        \includegraphics[scale=0.6, angle=0]{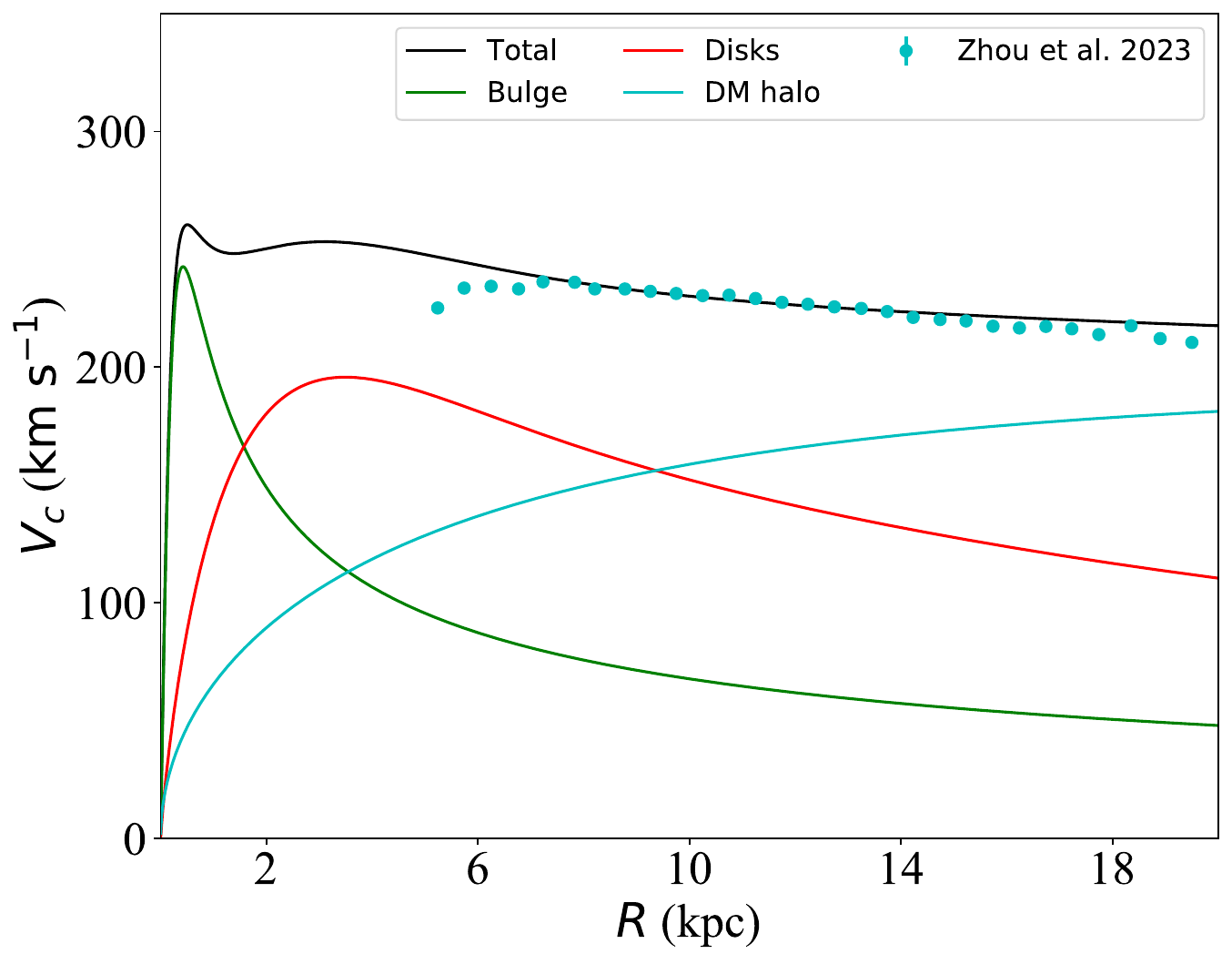}
    \end{center}
    \small{\textbf{Supplementary Figure 9}: Determination of $\rho_{s}$ and $r_{s}$ with the latest measurement of the Galactic rotation curve (cyan dots). The black line represents the best-fitted rotation curve,  which is the sum of contributions from the bulge (green line), the disks (red line,  $R_{d,\rm{thin}}=4\ \rm{kpc}$ as an example), and the DM halo (cyan line).}
\end{figure*}

\begin{figure*}
    \begin{center}
        \includegraphics[scale=0.085, angle=0]{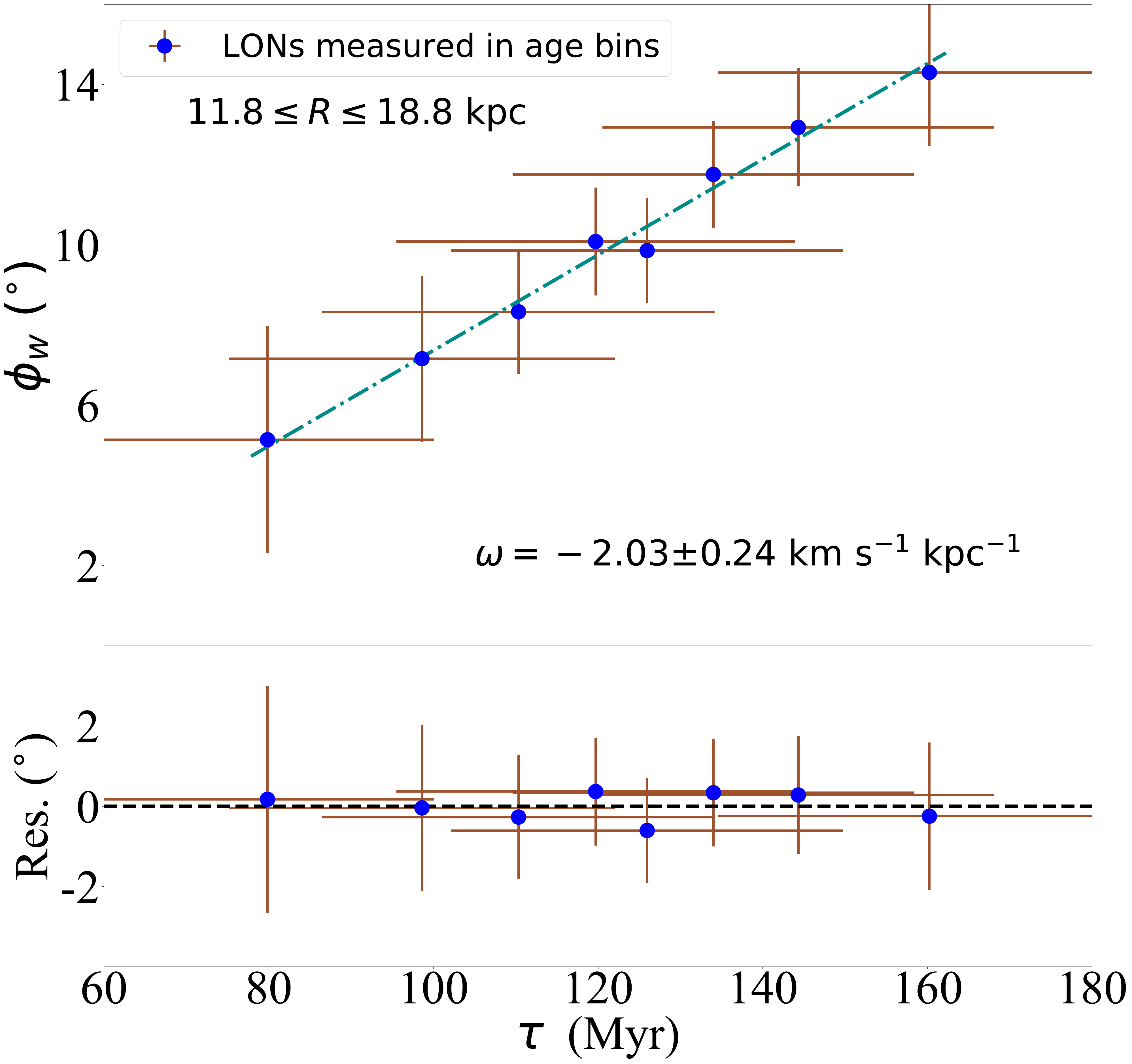}
        \includegraphics[scale=0.085, angle=0]{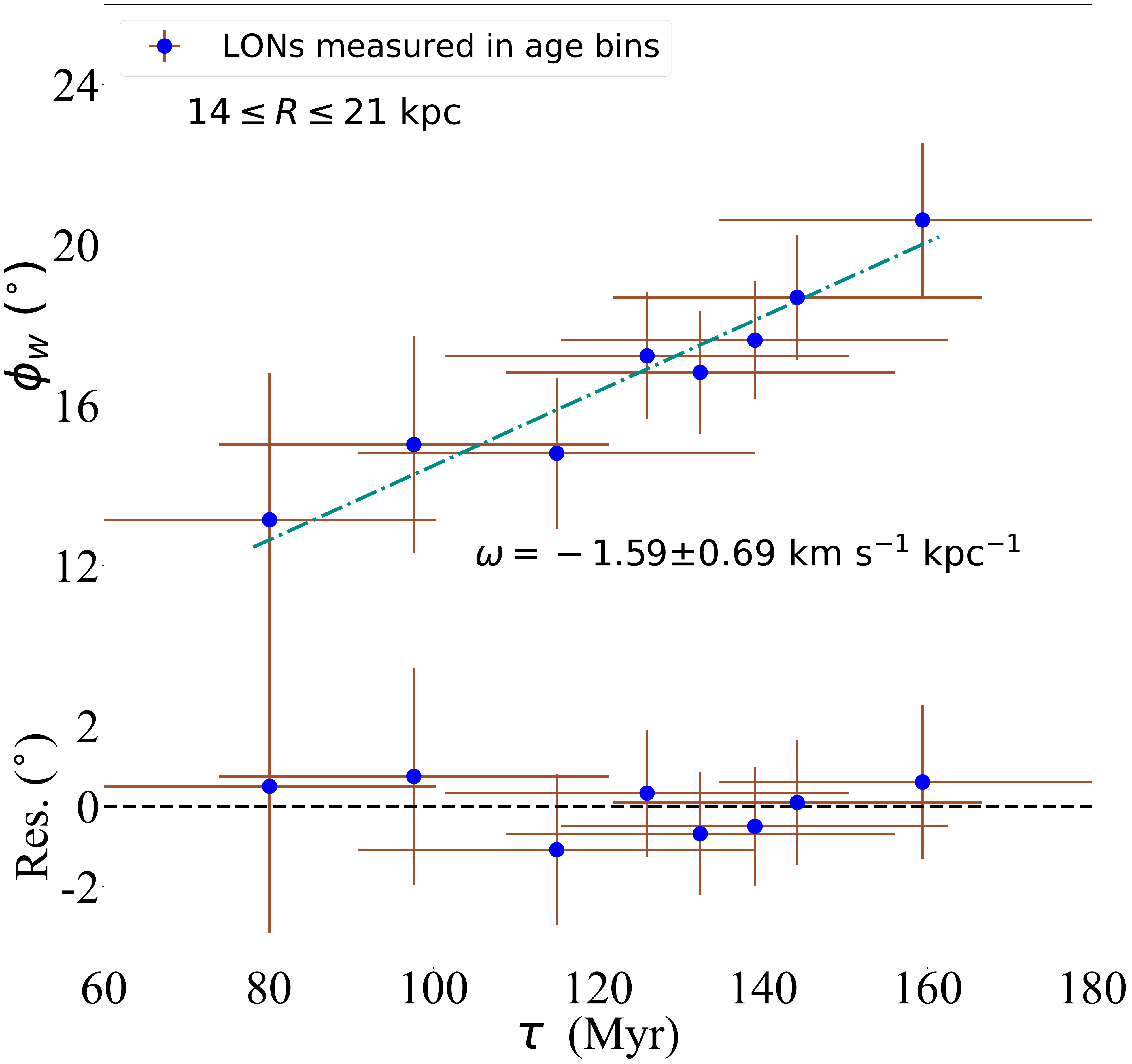}
        \includegraphics[scale=0.085, angle=0]{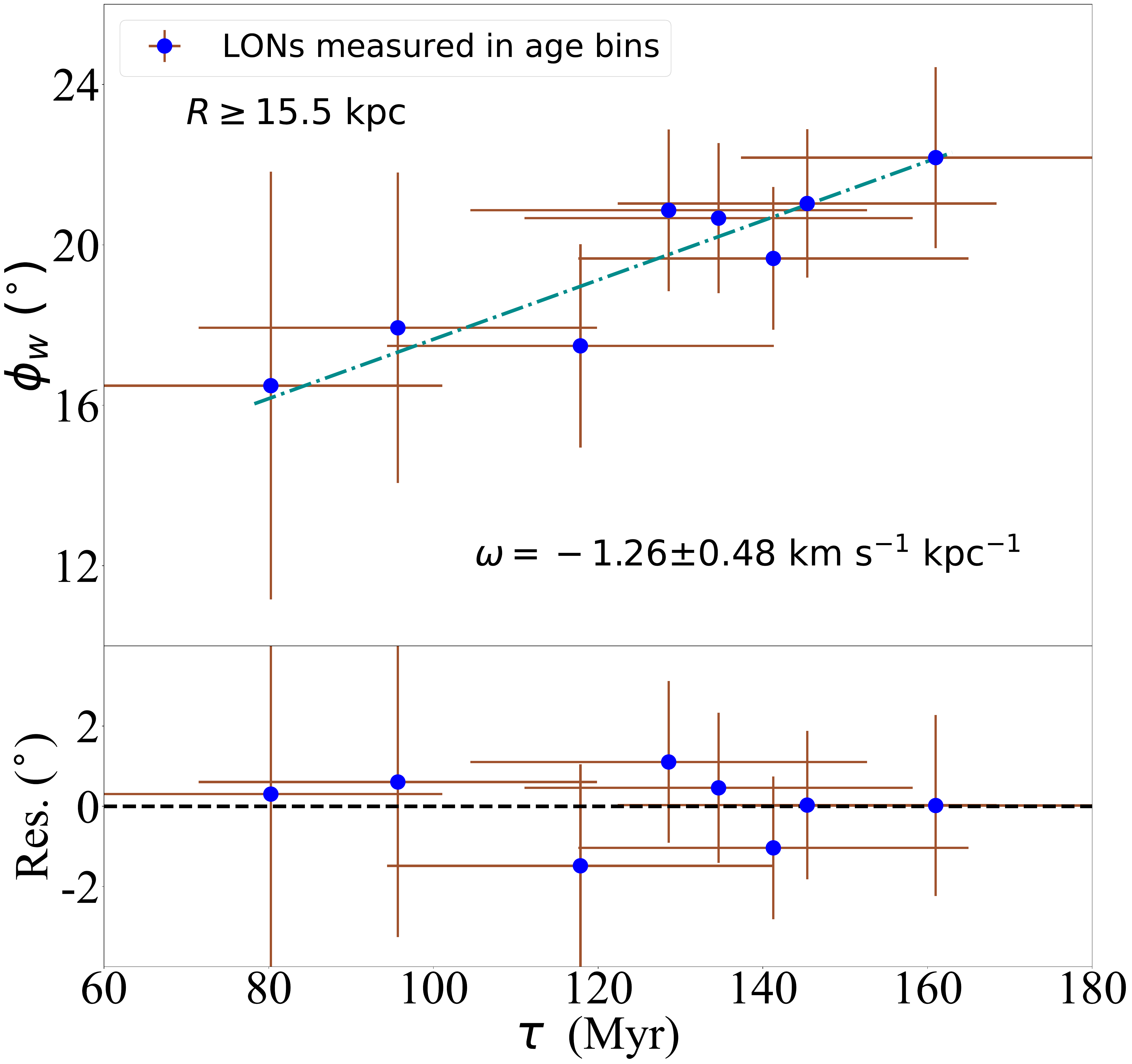}
    \end{center}
    \small{\textbf{Supplementary Figure 10}: The relation between the measured LON and age in three radial bins. Similar to Figure 1(d), but for three radial bins: $11.8 \le R \le 18.8$\,kpc (left panel), $14 \le R \le 21$\,kpc (middle panel), and $R \ge 15.4$\,kpc (right panel). All error bars represent $1\sigma$ confidence regions.}
\end{figure*}

\end{document}